\documentclass[reqno,draft]{amsart}
\usepackage{multicol, color}

\newcommand{\rem}[1]{}

\theoremstyle{plain}
\newtheorem{lemma}{Lemma}
\newtheorem{theorem}[lemma]{Theorem}
\newtheorem{corollary}[lemma]{Corollary}

\newtheorem{proposition}[lemma]{Proposition}

\theoremstyle{remark}
\newtheorem{remark}{Remark}

\newcommand*  {\Z} {{\mathbb Z}}

\def\aa{\alpha}
\def\a2{\alpha^2}

\def\dd{\delta}

\def\ss{\sigma}

\def\Dd{\Delta}

\def\vp{\varphi}
\def\gd{\nabla}

\begin{document}
\title[Rate of Convergence of 2D Leray-$\alpha$ to NSE]
{On The Rate Of Convergence Of  The Two-Dimensional $\alpha$-Models Of Turbulence To The
Navier-Stokes Equations}
\date{October 14, 2009}
\thanks{{\bf To appear in:}{\it Numerical Functional Analysis and Optimization}}

\author[Y.Cao]{Yanping Cao}
\address[Y.Cao]
{Department of Mathematics\\
University of California\\
Irvine, CA 92697-3875,USA}
\email{ycao@math.uci.edu}

\author[E.S. Titi]{Edriss S. Titi}
\address[E.S. Titi]
{Department of Mathematics \\
and  Department of Mechanical and  Aerospace Engineering \\
University of California \\
Irvine, CA  92697-3875, USA \\
{\bf ALSO}  \\
Department of Computer Science and Applied Mathematics \\
Weizmann Institute of Science  \\
Rehovot 76100, Israel}
\email{etiti@math.uci.edu and edriss.titi@weizmann.ac.il}

\begin{abstract}
Rates of convergence of solutions of various two-dimensional $\alpha-$regularization
models, subject to periodic boundary conditions, toward solutions of  the exact Navier-Stokes equations are given in the $L^\infty$-$L^2$ time-space norm, in terms of the regularization parameter $
\alpha$,
when $\alpha$ approaches zero. Furthermore, as a paradigm, error estimates for
the Galerkin approximation of the exact two-dimensional Leray-$\alpha$ model are also presented
in the $L^\infty$-$L^2$ time-space norm. Simply by the triangle inequality, one can reach the error
estimates of the solutions of Galerkin approximation of the $\alpha$-regularization models toward the exact solutions of the Navier-Stokes equations in the two-dimensional periodic boundary conditions case.
\end{abstract}

\maketitle



\section{Introduction} \label{sec-intro}
The closure problem of averaged quantities in turbulent flows has
been, for many years, an outstanding open challenge for turbulence
models. In the recent decade, various $\alpha$-regularization models
(Navier-Stokes-$\alpha$, Leray-$\alpha$, Modified Leray-$\alpha$,
Clark-$\alpha$ and simplified Bardina model) were introduced as
efficient subgrid scale models of the Navier-Stokes equations (NSE)
\cite{OBardina, Bardina,  CFHOTW, CFHOTWFluid,NSa, Leraya,FHT1,
FHT,HT, MLa,Layton} (see also \cite{OT} for an analytical study of a
mathematical generalization of the Navier-Stokes$-\alpha$ model). In
particular, it was shown in some of these papers that these
$\alpha$-models fit remarkably well with empirical experimental data
for a large range of huge Reynolds numbers. Moreover, these models
were implemented numerically by various groups \cite{BLTi, B, NSa,
CHMZ, GH,GH2, GKT,HN, HT,LL,LKT, LKTT, MKSM, NS}. Indeed, the
authors of \cite{LT} have pushed further this numerical analysis
point of view, concerning the $\alpha$-models, in their study of the
MHD-$\alpha$ model (see also \cite{Montgomery1, Montgomery2}). In
fact, there have been extensive analytical studies on the global
regularity of solutions and finite-dimensionality of global
attractor of these models, however, there is much less work on the
convergence, especially, the rate of convergence of solutions of
various $\alpha$-models toward the solutions of the exact NSE, when
the regularization parameter $\alpha$ approaches zero. The authors
of \cite{FHT} study the convergence of the three-dimensional (3D)
Navier-Stokes-$\alpha$ (NS-$\alpha$) model to the 3D NSE. To be more
specific, they show that there exists a subsequence of solutions
$u^{\alpha_j}$ of the 3D NS-$\alpha$ that converges to {\textit{one
of }}the Leray-Hopf weak solutions of the 3D NSE, with periodic
boundary conditions. Similar results are reported in \cite{LT}
concerning the MHD-$\alpha$ model. Later the authors of \cite{CTV,
VTC} show that the trajectory attractor of the Leray-$\alpha$ and
Navier-Stokes-$\alpha$, respectively, converges to the trajectory
attractor of the 3D Navier-Stokes system, as $\aa$ approaches zero.
Since the uniqueness theorem for global weak solutions (or global
existence of strong solutions) of the 3D NSE is not yet proved, the
studies mentioned above either consider convergence to a weak
solution or consider the convergence to the trajectory attractor of
weak solutions. Recently the authors of \cite{CGKTW} study the
convergence rate of the Navier-Stokes-$\alpha$ model and obtain a
mixed $L^1$-$L^2$ time-space norm for small initial data in
Besov-type function spaces in which global existence and uniqueness
of solutions for 3D NSE can be established with ``small" initial
data and external forcing. Similar results can also be derived by
applying the same techniques for the other $\alpha$-models under the
assumption of existence of strong solutions of the 3D NSE, e.g.,
when the initial value and external forcing are small
enough in the appropriate norms.\\\\
It is worth mentioning that inspired by the $\alpha$-regularization
models of turbulence similar regularization schemes were introduced
and implemented in \cite{CMT, CMT2}, in the context of the nonlinear
Schr\"{o}dinger equation, and in \cite{B} for the Leray-$\alpha$
regularization of the inviscid Burgers equation.\\\\
We mainly investigate in this paper the rates of convergence of four
$\alpha$-models (NS-$\alpha$ model, Leray-$\alpha$ model, Modified
Leray-$\alpha$ model and simplified Bardina model) in the
two-dimensional (2D) case, subject to periodic boundary conditions
on the periodic box $[0,L]^2$. Since unique strong solution is
proved to exist globally in time with any smooth enough initial data
in the 2D case, we will show upper bounds, in terms of $\alpha$, for
the difference between solutions of the 2D $\alpha$-models, $u^\aa$,
and solutions of the 2D NSE system, $u$, in the $L^2$-norm for any
time interval $[0,T]$. Specifically, we show that all the four
$\alpha$-models we include in this study have the same order of
convergence and error estimates, i.e., the $L^2$-norms of the
differences, $\|u-u^\alpha\|_2$, are of the order
$O\left((\frac{\alpha}{L})(\log(\frac{L}{\alpha}))^{1/2}\right)$, as
$ \frac{\alpha}{L}$ tends to zero. These results are presented in
detail in Section \ref{errorconv}. It is worth mentioning that the
Brezis-Gallouet inequality plays an essential role in our error
estimates in the 2D case; which in turn results in the logarithmic
factor. This logarithmic factor, however, is absent in the 3D case
which is treated in \cite{CGKTW}.
In a forthcoming paper, we will consider the rate of convergence of 3D Leray-$\alpha$ and NS-$\alpha$ models toward the adequate strong solution of the 3D exact NSE system, provided the latter exists, and compare the results to that of \cite{CGKTW}.\\\\
In Section \ref{errorGalerkin}, we consider the error estimates of
the Galerkin approximation solutions in the 2D case, i.e., we
estimate the difference between the solutions of the 2D
$\alpha$-model, $u^\aa$, and solutions of its corresponding
finite-dimensional Galerkin approximation system, $u_m^\aa$, where
$m\ge 1$ is the order of the truncation mode (dimension) of the
Galerkin system. We will study, as an example, the Leray-$\alpha$
model and present the detailed proof of the error estimates. One can
easily apply similar arguments to the other $\alpha$-regularization
models (NS-$\alpha$ model, Modified Leray-$\alpha$ model and
simplified Bardina model) and obtain similar error estimates for
these models. For the Leray-$\alpha$ model, the $L^2$-norm of the
difference, $\|u^\aa-u_m^\aa\|_2$, is of the order
$O\left(\frac{1}{\lambda_{m+1}L^2}(\log
(\lambda_{m+1}L^2))^{1/2}\right)$, under the assumption that
$\alpha$ is small such that $\alpha^2\le \frac{1}{\lambda_{m+1}}$,
where $\lambda_{m+1}$ is the $(m+1)-th$ eigenvalue of the Stokes
operator in the 2D case. Applying the triangle inequality, we get
error estimates concerning solutions of finite-dimensional Galerkin
system of the 2D Leray-$\alpha$ model, $u_m^\aa$, as an
approximation of the exact solution of the 2D NSE system, $u$.
Specifically, we show that the error in the $L^2$-norm of the
difference between the solution $u$ of the 2D NSE and $u^\aa_m$, the
solution of the Leray-$\alpha$ Galerkin system, $\|u-u_m^\aa\|_2$,
is of the order $O\left( \frac{1}{\lambda_{m+1}L^2}(\log
(\lambda_{m+1}L^2)^{1/2})\right)$, provided $\alpha$ is small enough
satisfying
$\alpha\le \frac{2\pi}{\lambda_{m+1}L}$.\\\\
Before we present the main results of the error estimates, we will first introduce in Section
\ref{sec-fun} some notations
and preliminaries that will be used throughout this paper. In Section \ref{errorapriori} we will
present all four $\alpha$-models in functional setting and establish \textit{a priori} estimates
for the solutions, by investigating the finite-dimensional Galerkin systems then passing to
the limit $m\rightarrow \infty$ by Aubin compactness theorem to obtain upper bounds for the exact solutions of the relevant models in certain norms.
The main results for the rate of convergence, in terms of the regularization parameter $\alpha$, of the $\alpha$-models in the 2D cases will be presented in Section \ref{errorconv}.
After establishing the rate of convergence of solutions of $\alpha$-regularization models
toward the solutions of the exact Navier-Stokes system, we will further
show in section \ref{errorGalerkin}  the error estimates of difference between solutions
of the 2D Leray-$\alpha$ model and solutions of its corresponding finite-dimensional Galerkin approximation system.
\section{Functional Setting and Preliminaries} \label{sec-fun}
In this section, we introduce some preliminary background material following the usual
notation used in the context of the mathematical theory of Navier-Stokes
equations (NSE) (see, e.g., \cite{CF}, \cite{SY}, \cite{Temam}, \cite{Temam2}).
\begin{enumerate}
\item We denote by $L^p$ and $H^m$ the usual Lebesgue and Sobolev spaces,
respectively.  And we denote by $|\cdot|$ and $(\cdot,\cdot)$ the $L^2$-norm
and $L^2$-inner product, respectively.
\item Let $\mathcal{F}$ be the set of trigonometric polynomials of two variables
with basic periodic domain $\Omega=[0,L]^2$ and spatial average zero, i.e., for every $\phi \in \mathcal{F}, \int_\Omega\phi(x)\:dx=0$.  We then set
$$
\mathcal{V}=\left\{\phi \in
(\mathcal{F})^2:\nabla\cdot\phi = 0 \right\}.
$$
We denote by $H$ and $V$ the closures of $\mathcal{V}$ in the $(L^2)^2$ and $(H^1)^2$ topologies, respectively. For $u\in H$ and $v\in V$, we denote by
\begin{equation*}
|u|^2=\int_\Omega|u(x)|^2\: dx \qquad \mbox{and} \qquad \|v\|^2=\int_\Omega|\gd v(x)|^2\:dx
\end{equation*}
the norms in $H$ and $V$, respectively.
We also note that by Rellich Lemma (see, e.g., \cite{Adams}), we have that $V$ is compactly embedded in $H.$
\item For any $s\ge0$, we denote by
$\dot{H}^s$ the closure of $\mathcal{F}$ with respect to the $H^s(\Omega)$ topology. Hence, for any
$u\in \dot{H}^s$, we can write the Fourier expansion
\begin{equation*}
\hskip-.8in
u(x)=\underset{ k \in \Z^2_0 }{\Sigma}\hat{u}_ke^{ i2\pi \frac{k\cdot x}{L}},
\end{equation*}
where the Fourier coefficients $\hat{u}_k$ satisfy the reality condition $\hat{u}_{-k}=\hat{u}^*_k$, $\forall k \in \Z^2_0:=\Z^2 \backslash \{0,0\}.$\\
We define the norm on this space as
\begin{equation*}
\hskip-.8in
\| u\|_{\dot{H}^s}^2=(\frac{2\pi}{L})^{2(s-1)}\underset{k\in \Z^2_0} {\Sigma}(1+|k|^2)^s |\hat{u}_k|^2.
\end{equation*}
\item We denote by $P_\ss:L^2 \rightarrow H$ the Helmholtz-Leray orthogonal
projection operator, and by $A = -P_\ss\Dd$ the Stokes operator, subject to
periodic boundary conditions, with domain $D(A) = (H^2(\Omega))^2\cap V$.  We
note that in the space-periodic case
\begin{equation*}
Au = -P_\ss\Dd u = -\Dd u, \hspace{.5cm} \mbox{for all }u \in D(A).
\end{equation*}
The operator $A^{-1}$ is a self-adjoint positive definite compact operator
from $H$ into $H$ (see, e.g., \cite{CF}, \cite{Temam}). We denote by
$0 < (\frac{2\pi}{L})^2= \lambda_1 \leq \lambda_2 \leq \dots \dots$ the
eigenvalues of $A$ in the 2D case, repeated according to their multiplicities.  It is well
known that in two dimensions the eigenvalues of the operator $A$ satisfy
the Weyl's type formula (see e.g., \cite{CF}, \cite{Temam2}), namely, there
exists a dimensionless constant $ c_0 > 0$ such that
\begin{equation}\label{asymptotic}
\dfrac{j}{c_0} \leq \dfrac{\lambda_j}{\lambda_1} \leq c_0 j, \hspace{.5 cm}\mbox{for } j = 1,2,\dots
\end{equation}
We also observe that in the periodic case,  $D(A^{n/2}) = (H^n(\Omega))^2\cap H$, for $n>0$.
In particular, one can show that
$V=D(A^{1/2})$ (see, e.g., \cite{CF}, \cite{Temam}).
\item  For every $w\in V$, we have the Poincar\'{e} inequality
\begin{equation}
\hskip-.8in
\lambda_1|w|^2\le \|w\|^2.                                                      \label{E1}
\end{equation}
Moreover, one can easily show that there is a dimensionless constant $c>0$, such that
\begin{equation}\label{A-norm}
c|Aw| \le \|w\|_{H^2} \le c^{-1}|Aw|  \qquad \mbox{for every } w\in D(A),
\end{equation}
and, by virtue of Poincar\'{e} inequality,
\begin{equation}{\label{H1-norm}}
c|A^{1/2} w| \le \|w\|_{H^1} \le c^{-1} |A^{1/2}w| \qquad \mbox{for every } w\in V.
\end{equation}
Hereafter, $c$ will always denote a generic dimensionless constant.\\
Notice that, thanks to (\ref{H1-norm}), the norm of $V$ is equivalent
to the usual $H^1$ norm.
\item Let $\{w_j\}_{j=1}^\infty$ be an orthonormal basis of $H$ consisting of eigenfunctions of the operator $A$. Denote by $H_m=\mbox{span}\{w_1,w_2, ...,w_m\}$, for $m \ge1$ and let $P_m$ be the $L^2$-orthogonal projection from $H$ onto $H_m$, then it is easy to see that
\begin{eqnarray}
\hskip-.8in
&& |(I-P_m)\phi |^2 \le \frac{1}{\lambda_{m+1}} \| \phi\|^2,\quad \mbox{for all } \phi\in V, \label{pc1}\\
&& \| (I-P_m)\phi\|^2\le \frac{1}{\lambda_{m+1}} |A\phi|^2 ,\quad \mbox{for all } \phi \in D(A). \label{pc2}
\end{eqnarray}
Moreover, one can also easily show that
\begin{equation}
\hskip-.8in
|A P_m \phi|^2 \le \lambda_{m} \|P_m \phi\|^2,  \quad \mbox{for all } \phi \in D(A). \label{log}
\end{equation}
\item We recall the following 2D interpolation and Gagliardo-Nirenberg-Ladyzhenskaya
inequality (see, e.g., \cite{Adams}, \cite{CF}, \cite{Temam}) :
\begin{equation}\label{sobolev}
\|\phi\|_{L^4}\leq c\|\phi\|_{L^2}^{1/2}\|\phi\|_{H^1}^{1/2}.
\end{equation}
\item For every $w_1, w_2 \in \mathcal{V}$, we define the bilinear operators
\begin{eqnarray}
&&B(w_1,w_2) = P_\ss((w_1\cdot\nabla)w_2),  \label{B}      \\
&&\tilde{B}(w_1,w_2)=-P_{\sigma}(w_1\times(\gd \times w_2)).    \label{tildeB}
\end{eqnarray}
\end{enumerate}
In the following lemma, we will list certain relevant inequalities and properties of $B$ (see, e.g., \cite{CF},
\cite{Temam}, \cite{Temam2}) and of $\tilde{B}$
(see \cite{FHT}).
\begin{lemma}\label{bilinearprop}
The bilinear operator B defined in (\ref{B}) satisfies the following:
\begin{enumerate}
\item $B$ can be extended as a continuous bilinear map $B:V \times V \rightarrow V'$,
where $V'$ is the dual space of $V$.  \\In particular,
the bilinear operator $B$ satisfies the following inequalities:
\begin{eqnarray}
&&\hskip-.6in
|\langle B(u,v),w\rangle_{V'}| \leq c|u|^{\frac{1}{2}}\|u\|^{\frac{1}{2}}\|v\| |w|^{\frac{1}{2}}\|w\|^{\frac{1}{2}},
\: \mbox{for all } \: u, v,w \in V,\label{b-uvw} \\
&&\hskip-.6in
|\left(B(u,v),w\right) |\le c\|u\|_{\infty} \|v\| |w|, \quad \mbox{for all } u\in D(A),\: v\in V,\: w\in H,\label{b-infty}  \\
&& \hskip-.6in
|\left(B(u,v),w\right)| \le c|u| \|\gd v\|_{\infty} |w|, \quad \mbox{ for all } u \in H,\: v \in D(A^{3/2}),\: w\in H,\label{b-infty2} \\
&& \hskip-.6in
|\langle B(u,v),w\rangle_{(D(A))'}| \le c |u| \|v\| \| w\|_{\infty}, \:\mbox{ for all } u\in H, \: v\in V,\ w\in D(A).\label{b-infty3}
\end{eqnarray}
Moreover, for every $w_1,w_2,w_3 \in V$, we have
\begin{equation}\label{bilinear}
\hskip-.8in
\langle B(w_1,w_2),w_3\rangle_{V'} = -\langle B(w_1,w_3),w_2\rangle_{V'},
\end{equation}
and in particular,
\begin{equation}\label{b-uvv}
\hskip-.8in
\langle B(w_1,w_2),w_2\rangle_{V'}= 0.
\end{equation}
\item In the 2D periodic boundary condition case, we have
\begin{equation}\label{b-uuau}
\hskip-.8in
(B(\vp, \vp), A\vp)= 0,
\end{equation}
for every $\varphi \in D(A)$.
\item $\tilde{B}$ can also be extended as a continuous bilinear map $\tilde{B}: V\times V \rightarrow V'$. Furthermore, for every $w_1, w_2, w_3 \in V$, the bilinear operator $\tilde{B}$ satisfies the following inequality:
\begin{equation*}
|\langle \tilde{B}(w_1,w_2),w_3\rangle_{V'}| \leq c|w_1|^{1/2}\|w_1\|^{1/2}\|w_2\| |w_3|^{1/2}\|w_3\|^{1/2}.
\end{equation*}
Moreover, for every $w_1,w_2,w_3\in V$, we have
\begin{equation}
\langle \tilde{B}(w_1,w_2),w_3\rangle_{V'}
=\langle B(w_1,w_2),w_3\rangle_{V'}-\langle B(w_3,w_2),w_1\rangle_{V'},         \label{tildeB-id}
\end{equation}
and consequently,
\begin{equation}
\hskip-.8in
\langle \tilde{B}( w_1, w_2), w_1\rangle_{V'}=0.            \label{tildeB-zero}
\end{equation}
\end{enumerate}
\end{lemma}
\begin{lemma}\label{B-lemma}
For every $u \in D(A)$ and $w\in V$, we have
\begin{equation*}
\hskip-.2in
|(B(w,u),Au)|  \le c \|w\| \|u\| |Au|.
\end{equation*}
\end{lemma}
\begin{proof}
Let $w\in \mathcal{V}$ and $u \in D(A)$. Then
\begin{eqnarray*}
\hskip-.4in
(B(w,u),Au)
&=&- \overset{2}{\underset{l,j,m=1}{\Sigma} }\int_{\Omega} w_j(\frac{\partial}{\partial x_j} u_l)
(\frac{\partial^2}{\partial x_m^2}u_l)\:dx    \\
&=&\overset{2}{\underset{l,j,m=1}{\Sigma} } \left[ \int_{\Omega}
(\frac{\partial}{\partial x_m}w_j)(\frac{\partial}{\partial x_j} u_l)(\frac{\partial}{\partial x_m}u_l)\: dx
+\int_{\Omega}w_j(\frac{\partial }{\partial x_j}\frac{\partial}{\partial x_m}u_l)\frac{\partial}{\partial x_m}u_l)
\: dx \right].
\end{eqnarray*}
By relation (\ref{b-uvv}) the second term on the right-hand side above is zero. Therefore, by the Cauchy-Schwarz inequality
\begin{equation*}
\hskip-.8in
|(B(w,u),Au)| \le c\|w\| \| \gd u\|_{L^4}^2,
\end{equation*}
and by (\ref{sobolev}), (\ref{A-norm}) and (\ref{H1-norm}) we have
\begin{equation*}
\hskip-.8in
|(B(w,u),Au)| \le c \|w\| \|u\| |Au|.
\end{equation*}
From the above and the density of $\mathcal{V}$ in $V$ we conclude our lemma.
\end{proof}
Next, we state a two-dimensional periodic boundary condition version of the well-known Brezis-Gallouet inequality \cite{BG}. For the sake of completeness, we will present the proof of this version in the
Appendix.
\begin{proposition}\label{BG-inequality}
There exists a scale invariant constant $c> 0$ such that for every $\vp \in D(A)$,
\begin{equation*}
\hskip-.8in
\| \vp \|_{L^{\infty}} \le c \| \vp \| \left(  1+\log\left( \frac{L}{2\pi} \frac{ |A\vp|     }{\| \vp \|}\right)\right)^{1/2}.
\end{equation*}
\end{proposition}
\noindent Applying the Brezis-Gallouet inequality above to
(\ref{b-infty}) and (\ref{b-infty3}), we have the following
corollary (see also \cite{Titi1} for similar logarithmic
inequalities concerning the bilinear term).
\begin{corollary}\label{BG-corollary}
In the two-dimensional case, the bilinear operator $B$ satisfies the following inequalities:
\begin{eqnarray}
\hskip-.8in
&&|\left( B(u,v),w\right)| \le c\|u\| \|v\| |w| \left( 1+\log \frac{|Au|^2}{\|u\|^2\lambda_1}\right)^{1/2} \mbox{ for all } u\in D(A),\: v\in V,\: w\in H, \label{b-bg}\\
\hskip-.8in
&&|\left(B(u,v),w\right)| \le c|u|\|v\|\|w\|\left(1+\log \frac{|Aw|^2}{\|w\|^2\lambda_1}\right)^{1/2}, \mbox{ for all } u\in H,\: v\in V,\: w\in D(A). \label{b-bg2}
\end{eqnarray}
where $\lambda_1=(\frac{2\pi}{L})^2$ is the first eigenvalue of the Stokes operator $A$.\\
\end{corollary}
\section{\textit{A Priori} Estimates of the 2D Navier-Stokes Equations and the $\alpha$-Regularization Models}\label{errorapriori}
In this section, we will establish \textit{a priori} estimates for solutions of the 2D NSE and the 2D $\alpha$-regularization models.
These results will be useful for the error estimates in the next two sections.
\subsection{\textit{A priori} estimates for the 2D Navier-Stokes equations}\label{subsecNSEapriori}
We recall that the two dimensional (2D) NSE is equivalent (see, e.g.,
\cite{CF}, \cite{Temam}) to the functional evolution equation in the Hilbert space H
\begin{eqnarray}
\frac{du}{dt} +\nu Au+B(u,u) &=&f,                 \label{NSE}   \\
u(0)&=&u_0.         \nonumber
\end{eqnarray}
The corresponding Galerkin system of the 2D NSE is given below as a system of ordinary differential equations in the space $H_m$, defined in Section \ref{sec-fun}:
\begin{eqnarray}
\hskip-.8in
\frac{d u_m}{dt}+\nu Au_m+P_mB(u_m,u_m) &=& f_m,   \label{NSEGal}\\
u_m(0)=u_{0m}. \nonumber
\end{eqnarray}
where $u_{0m}=P_m\:u_0,\:f_m=P_m\:f$.\\\\
The proof of existence and uniqueness of solutions of the above 2D NSE system (\ref{NSE}), subject to periodic boundary
conditions, can be established by applying Galerkin approximation procedure \cite{CF, Temam, Temam2}. The idea is to establish \textit{a priori} estimates
for the solutions of the finite-dimensional system (\ref{NSEGal}) and then by applying Aubin compactness theorem (see also \cite{CF, Temam, Temam2}) one can extract a subsequence that converges to the unique solution of the NSE system (\ref{NSE}). The details of the proof are textbook material and will not be presented here. In the following
proposition we will establish \textit{a priori} estimates for solutions of the Galerkin system ({\ref{NSEGal}) that we will use later.
\begin{proposition}\label{NSE-estimate}
Let $u_0 \in H,T>0$ and $f \in L^2((0,T);H)$. Let $u_m$ be the solution of the Galerkin approximation of the 2D NSE, system (\ref{NSEGal}), over the interval $[0,T]$ with initial data $u_{0m}$ for a given $m\ge 1$, then
\begin{equation}
\underset{0\le t\le T}{\sup} \left(  |u_m(t)|^2+\nu \int_0^t \|u_m(s)\|^2 \:ds    \right) \le K_0^2(\nu, u_0, f, L, T), \label{K-1}
\end{equation}
where $K_0^2:=|u_0|^2+\frac{1}{\nu\lambda_1} \int_0^T|f(s)|^2\: ds$ and $\lambda_1=(\frac{2\pi}{L})^2$.
\end{proposition}
\begin{proof}
We take the inner product of the first equation in (\ref{NSEGal}) with $u_m$ to obtain
\begin{equation*}
\hskip-.8in
\frac{1}{2}\frac{d}{dt}|u_m|^2+\nu \|u_m\|^2=(P_mf,u_m)\le |f| |u_m|.
\end{equation*}
By Poincar\'{e} inequality (\ref{E1}) we have
\begin{equation*}
\hskip-.8in
\frac{1}{2}\frac{d}{dt}|u_m|^2+\nu \|u_m\|^2
\le \frac{|f| \|u_m\|}{\lambda_1^{1/2}} \le\frac{ |f|^2}{2\nu\lambda_1} +\frac{\nu}{2} \|u_m\|^2.
\end{equation*}
Hence,
\begin{equation*}
\hskip-.8in
\frac{d}{dt}|u_m|^2 +\nu \|u_m\|^2 \le \frac{ |f|^2}{\nu\lambda_1},
\end{equation*}
and by integration we have
\begin{equation*}
\hskip-.8in
|u_m(t)|^2+\nu \int_0^t \|u_m(s)\|^2 \: ds \le |u_0|^2+ \frac{1}{\nu\lambda_1} \int_0^t |f(s)|^2 \: ds \quad
\mbox{ for all } t\in [0,T],
\end{equation*}
which concludes the proof.
\end{proof}
\begin{remark}
One needs to further establish estimates on $\frac{du_m}{dt}$ and then, thanks to Aubin compactness theorem \cite{CF, Temam}, one can extract a subsequence $u_{m'}$,
that converges to the unique solution $u$ of system (\ref{NSE}), as $m'\rightarrow \infty$. Following the standard procedure as in \cite{CF, Temam}, one can show that the solution of the NSE (\ref{NSE}) satisfies the same \textit{a priori} estimates, namely, let
$u$ be the solution of the system (\ref{NSE}) over the interval $[0,T]$ with initial data $u_0 \in H$, then
\begin{equation}
\underset{0\le t\le T}{\sup} \left(  |u(t)|^2+\nu \int_0^t \|u(s)\|^2 \:ds    \right) \le K_0^2(\nu, u_0, f, L, T). \label{NSEapriori}
\end{equation}
\end{remark}
\subsection{\textit{A priori} estimates for the 2D Leray-$\alpha$ model}\label{subsecLAapriori}
In this subsection,we will establish \textit{a priori} estimates for solutions of the 2D Leray-$\alpha$
regularization model. The Leray-$\alpha$ model was introduced and analyzed in \cite{Leraya} and implemented computationally \cite{Leraya, GH,GH2, GKT, LKT} in the context of subgrid scale models of 3D turbulence (see also
\cite{B, HN} and \cite{LKT} for 2D computations with the Leray-$\alpha$ model). The Leray-$\alpha$ model
was inspired by the Navier-Stokes-$\alpha$ model (also known as viscous Camassa-Holm or Lagrangian-averaged-Navier-Stokes-$\alpha$ model) \cite{CFHOTW,CFHOTWFluid,  NSa, FHT1, FHT},
and it happened
to fit as a member of the general family of regularizations introduced in the seminal work of
Leray \cite{Leray} in the context of establishing the existence of solutions for the 2D and 3D NSE.\\\\
The Leray-$\alpha$ regularization model of the NSE is given by the following
functional evolution system in the space $H$:
\begin{eqnarray}
\frac{dv^\aa}{dt}+\nu Av^\aa+B(u^\aa,v^\aa) &=& f,          \label{LA}   \\
u^\aa+\alpha^2 Au^\aa  &=& v^\aa,  \nonumber   \\
u^\aa(0) &=&  u_0.               \nonumber
\end{eqnarray}
Observe that when the regularization parameter $\alpha=0$ one recovers the exact NSE system (\ref{NSE}).\\\\
The corresponding Galerkin system of the 2D Leray-$\alpha$ model is given below as a system
of ordinary differential equations in the space $H_m$:
\begin{eqnarray}
\frac{dv^\aa_m}{dt}+\nu Av^\aa_m+P_m B(u^\aa_m, v^\aa_m)=f_m,  \label{LAGalerkin} \\
u^\aa_m+\alpha^2Au^\aa_m=v_m^\aa,  \nonumber \\
u_m^\aa(0)=u_{0m}, \nonumber
\end{eqnarray}
where $u_{0m}=P_m \:u_0,\: f_m=P_m\: f$.\\\\
The proof of existence and uniqueness of solution of the above Leray-$\alpha$ system (\ref{LA}) and the other three $\alpha$-models we cover in this section,
subject to periodic boundary
conditions, can be established by applying Galerkin approximation procedure. One can follow similar steps as those for the NSE
\cite{CF, Temam, Temam2} to show the proof. The proof is not the heart of this paper and we will omit the details and only establish \textit{a priori} estimates for
solutions of the Galerkin system (\ref{LAGalerkin}). To be concise and focus on the essential matter of this paper, in this and the following subsections of \textit{a priori} estimates for the 2D $\alpha$-models,
we will simply skip the details of the proof and will not restate these comments.
Interested readers can refer to the relevant literature in \cite{CF, Temam, Temam2} and references therein to fill in the gap.
\begin{proposition}\label{LA-estimate}
Let $u_0 \in D(A), T>0$ and $f\in L^2((0,T);H)$. Let $u^\aa_m$ be the solution of the system (\ref{LAGalerkin}) over the interval $[0,T]$ with initial
data $u_{0m}$ for a given $m\ge1$, then
\begin{equation*}
\hskip-.8in
\underset{0\le t\le T}{\sup} \left[   \|u^\aa_m(t)\|^2 +\alpha^2 |Au^\aa_m(t)|^2 +\nu \int_0^t (|Au^\aa_m(s)|^2+\alpha^2 |A^{3/2}u^\aa_m(s)|^2) ds    \right] \le {\tilde{K}}_0^2
\end{equation*}
where ${\tilde{K}}_0^2:=\|u_{0}\|^2+ \alpha^2|Au_0|^2 +\frac{1}{\nu}\int_0^T  |f(s)|^2 \: ds$.
\end{proposition}
\begin{proof}
We take the inner product of the first equation in system (\ref{LAGalerkin}) with $Au^\aa_m$ to obtain
\begin{equation*}
\hskip-.2in
\frac{1}{2}\frac{d}{dt} ( \|u^\aa_m\|^2+\alpha^2 |Au^\aa_m|^2)+ \nu (|Au^\aa_m|^2+ \alpha^2 |A^{3/2}u^\aa_m|^2)+(B(u^\aa_m,v^\aa_m),Au^\aa_m)=(f, Au^\aa_m).
\end{equation*}
By (\ref{b-uvv}) and (\ref{b-uuau}) one has
\begin{equation*}
\hskip-.8in
(B(u^\aa_m,v^\aa_m),Au^\aa_m)=(B(u^\aa_m,u^\aa_m),Au^\aa_m)+\alpha^2(B(u^\aa_m,Au^\aa_m),Au^\aa_m)=0.
\end{equation*}
Consequently, from the above we have
\begin{eqnarray*}
\frac{1}{2} \frac{d}{dt} (\| u^\aa_m\|^2+\alpha^2 |Au^\aa_m|^2) +\nu (|Au^\aa_m|^2+\alpha^2 |A^{3/2}u^\aa_m|^2)
&=&(f, Au^\aa_m) \le |f| |Au^\aa_m|           \\
&\le&\frac{   |f|^2  }{2\nu}+\frac{\nu}{2}|Au^\aa_m|^2.
\end{eqnarray*}
Hence
\begin{equation*}
\hskip-.6in
\frac{d}{dt}(\|u^\aa_m\|^2+\alpha^2 |Au^\aa_m|^2)+\nu ( |Au^\aa_m|^2+ \alpha^2 |A^{3/2} u^\aa_m|^2) \le \frac{  |f|^2    }{ \nu },
\end{equation*}
and by integration over the interval $(0,t)$, for $t\in [0,T]$ we have
\begin{eqnarray*}
\|u^\aa_m(t)\|^2+\alpha^2 |Au^\aa_m(t)|^2
&+&\nu \int_0^t ( |Au^\aa_m(s)|^2+\alpha^2|A^{3/2}u^\aa_m(s)|^2)\: ds \\
&\le& \frac{1}{\nu} \int_0^T |f(s)|^2 \:ds+\|u_{0m}\|^2+\alpha^2 |Au_{0m}|\\
&\le& \frac{1}{\nu}\int_0^T|f(s)|^2\:ds +\|u_0\|^2+\alpha^2 |Au_0|,
\end{eqnarray*}
which concludes our proof.
\end{proof}
\begin{remark}
Similar to the 2D NSE, one needs to further establish estimates on $\frac{du^\aa_m}{dt}$ and then, thanks to Aubin compactness theorem \cite{CF, Temam},  one can extract a subsequence $u^\aa_{m'}$
that converges to the unique solution $u^\aa$ of system (\ref{LA}), as $m'\rightarrow \infty$. As a result, one can also prove that the solution of the Leray-$\alpha$ system (\ref{LA}) satisfies the same \textit{a priori} estimates as
in Proposition \ref{LA-estimate}, namely, let $u^
\aa$ be the solution of system (\ref{LA}) over the interval $[0,T]$ with initial data $u_0\in D(A)$,  then
\begin{equation}
\hskip-.8in
\underset{0\le t\le T}{\sup} \left[   \|u^\aa(t)\|^2 +\alpha^2 |Au^\aa(t)|^2 +\nu \int_0^t (|Au^\aa(s)|^2+\alpha^2 |A^{3/2}u^
\aa(s)|^2) ds    \right] \le {\tilde{K}}_0^2. \label{LA-apriori}
\end{equation}
We emphasize here again that the details of the proof of the above results are omitted. In the following subsections of \textit{a priori} estimates of the NS-$\alpha$, Modified Leray-$\alpha$ and simplified Bardina models, we will also skip the details and will not restate the remark.
\end{remark}
\subsection{\textit{A priori} estimates for the 2D NS-$\alpha$ model}\label{subsecNSaapriori}
In this subsection, we will establish \textit{a priori} estimates for the 2D NS-$\alpha$ regularization
model of the 2D NSE.
The
NS-$\alpha$ model (also known as the viscous Camassa-Holm or Lagragian-averaged-Navier-Stokes-
$\alpha$ model) was introduced and analyzed in
\cite{CFHOTW, CFHOTWFluid, NSa, FHT1, FHT, HT}, which was also the first
of the family of the $\alpha$-models. In addition to the remarkable match of explicit solutions of the
3D NS-$\alpha$ model to experimental data, in the channels and pipes, for a wide range of
huge Reynold numbers \cite{CFHOTW, CFHOTWFluid, NSa}, the validity of NS-$\alpha$ model as a subgrid scale turbulence model was tested numerically in \cite{NSa, CHMZ, GKT,HN, LKTT, MKSM, NS}. \\\\
The NS-$\alpha$ regularization model of the NSE is given by the following functional
evolution system in the space H:
\begin{eqnarray}
\frac{dv^\aa}{dt}+\nu Av^\aa+\tilde{B}(u^\aa,v^\aa) &=& f,    \label{NSa} \\
u^\aa+\alpha^2 Au^\aa &=& v^\aa,       \nonumber \\
u^\aa(0)&=&u_0,                                                              \nonumber
\end{eqnarray}
where $\tilde{B}$ is given by (\ref{tildeB}) and Lemma \ref{bilinearprop}.\\
Observe that when the regularization parameter $\alpha=0$ one recovers the exact NSE system (\ref{NSE}).\\\\
The corresponding Galerkin approximation system of the 2D NS-$\alpha$ model is given below as a system of ordinary differential equations in the space $H_m$:
\begin{eqnarray}
\hskip-.8in
\frac{dv^\aa_m}{dt}+\nu Av^\aa_m +P_m \tilde{B}(u^\aa_m,v^\aa_m)&=& f_m, \label{NSaGalerkin}   \\
u^\aa_m+\alpha^2Au^\aa_m &=& v^\aa_m,   \nonumber    \\
u^\aa_m(0)&=&u_{0m},   \nonumber
\end{eqnarray}
where $u_{0m}=P_m\:u_0,\: f_m=P_m\:f$.\\\\
The proof of existence and uniqueness of solutions of the NS-$\alpha$ system (\ref{NSa}), subject to periodic boundary
conditions, is similar to that of the $3D$ case as it is presented in \cite{FHT}. In the next proposition we will omit the details and only show the \textit{a priori} estimates for solutions of system (\ref{NSaGalerkin}).
\begin{proposition}\label{NSa-estimate}
Let $u_0\in D(A),T>0$ and $f \in L^2((0,T);H)$. Let $u^\aa_m$ be the solution of the Galerkin system of 2D NS-$\alpha$ (\ref{NSaGalerkin}) over the interval $[0,T]$
with initial data $u_{0m}$ for a given $m\ge 1$, then
\begin{equation}
\hskip-.8in
\underset{0\le t\le T}{\sup}\left[ |u^\aa_m(t)|^2+\alpha^2 \|u^\aa_m(t)\|^2 +\nu \int_0^t (\|u^\aa_m(s)\|^2
+\alpha^2 |Au^\aa_m(s)|^2) \: ds \right]\le \tilde{\tilde{K}}_{01}^2,               \label{tildeK0}
\end{equation}
where $\tilde{\tilde{K}}_{01}^2:=|u_0|^2+\alpha^2\|u_0\|^2+\frac{1}{\nu\lambda_1} \int_0^T |f(s)|^2\: ds$, which is independent of $m$, and
$\lambda_1=(\frac{2\pi}{L})^2$.\\
Moreover,
\begin{equation}
\hskip-.8in
\underset{0\le t\le T}{\sup}\left[  \|u^\aa_m(t)\|^2+\alpha^2 |Au^\aa_m(t)|^2 +\nu \int_0^t (|Au^\aa_m(s)|^2+\alpha^2 |A^{3/2}u^\aa_m(s)|^2)\:ds        \right]   \le \tilde{\tilde{K}}_{02}^2,       \label{ttildeK0}
\end{equation}
where $\tilde{\tilde{K}}_{02}^2:= \|u_0\|^2+\alpha^2 |Au_0|^2+\frac{1}{\nu}\int_0^T |f(s)|^2\:ds+\frac{c}{\nu^2}
\tilde{\tilde{K}}_{00}^2\tilde{\tilde{K}}_{01}^2$, and $\tilde{\tilde{K}}_{00}$ is a constant depending on $|Au_0|,\nu,T, f$ and $\tilde{\tilde{K}}_{01}$, which is give explicitly in (\ref{tildeK01}).
\end{proposition}
\begin{proof}
By taking the inner product of the first equation in (\ref{NSaGalerkin}) with $u^\aa_m$ and using (\ref{tildeB-zero}) we obtain
\begin{eqnarray*}
\hskip-.4in
\frac{1}{2}\frac{d}{dt}(|u^\aa_m|^2+\alpha^2\|u^\aa_m\|^2)+\nu(\|u^\aa_m\|^2+\alpha^2 |Au^\aa_m|^2)
&=& (P_mf,u^\aa_m)\le|f||u^\aa_m| \le \frac{|f|\|u^\aa_m\|}{\lambda_1^{1/2}}            \\
&\le& \frac{|f|^2}{2\nu \lambda_1}+\frac{\nu \|u^\aa_m\|^2}{2},
\end{eqnarray*}
where in the last two steps we used Poincar\'{e} inequality (\ref{E1}) and Young's inequality.\\
As a result we have
\begin{equation*}
\hskip-.8in
\frac{d}{dt}(|u^\aa_m|^2+\alpha^2 \|u^\aa_m\|^2)+\nu(\|u^\aa_m\|^2+\alpha^2 |Au^\aa_m|^2) \le \frac{|f|^2}{\nu\lambda_1}.
\end{equation*}
Integrating the above in time we obtain (\ref{tildeK0}).\\\\
Next, we prove (\ref{ttildeK0}). We take the inner product of the equation in (\ref{NSaGalerkin})
with $Au^\aa_m$ and follow similar steps as above to obtain
\begin{eqnarray}
\hskip-.2in
\frac{1}{2}\frac{d}{dt} ( \|u^\aa_m\|^2+\alpha^2 |Au^\aa_m|^2)
&+& \nu (|Au^\aa_m|^2+\alpha^2 |A^{3/2}u^\aa_m|^2)+(\tilde{B}(u^\aa_m,v^\aa_m),Au^\aa_m) \nonumber \\
&=&(P_mf,Au^\aa_m)
\le\frac{|f|^2}{2\nu}+\frac{\nu}{2}|Au^\aa_m|^2.      \label{pro-1}
\end{eqnarray}
Observe that
\begin{equation}
\hskip-.0in
(\tilde{B}(u^\aa_m,v^\aa_m),Au^\aa_m)=(B(u^\aa_m,v^\aa_m),Au^\aa_m)-(B(Au^\aa_m,v^\aa_m),u^\aa_m)=\alpha^2(B(Au^\aa_m,u^\aa_m),Au^\aa_m),  \label{pro-2}
\end{equation}
where we use in the above (\ref{tildeB-id}), (\ref{bilinear}), (\ref{b-uvv}), (\ref{b-uuau})
and the relation $v^\aa_m=u^\aa_m+\alpha^2Au^\aa_m$.\\
From (\ref{pro-2}) and Lemma \ref{B-lemma} we have
\begin{equation*}
|(\tilde{B}(u^\aa_m,v^\aa_m),Au^\aa_m)|\le c\alpha^2|A^{3/2}u^\aa_m| \|u^\aa_m\||Au^\aa_m| \le \frac{\nu}{2}\alpha^2|A^{3/2}u^\aa_m|^2+\frac{c\alpha^2}{\nu}\|u^\aa_m\|^2|Au^\aa_m|^2.
\end{equation*}
From the above and (\ref{pro-1}) we obtain
\begin{eqnarray}
\hskip-.6in
\frac{d}{dt}( \|u^\aa_m\|^2+\alpha^2|Au^\aa_m|^2)
&+&\nu(|Au^\aa_m|^2+\alpha^2|A^{3/2}u^\aa_m|^2) \nonumber \\
&\le& \frac{|f|^2}{\nu}+\frac{c\alpha^2}{\nu}\|u^\aa_m\|^2|Au^\aa_m|^2        \nonumber \\
&\le& \frac{|f|^2}{\nu}+\frac{c}{\nu}(\|u^\aa_m\|^2+\alpha^2|Au^\aa_m|^2)^2.                \label{pro-3}
\end{eqnarray}
Therefore, by Gronwall's inequality we have
\begin{eqnarray*}
\hskip-.2in
\|u^\aa_m(t)\|^2+\alpha^2 |Au^\aa_m(t)|^2
&\le& e^{\frac{c}{\nu}\int_0^t(\|u^\aa_m(s)\|^2+\alpha^2 |Au^\aa_m(s)|^2)ds}(\|u_0\|^2 +\alpha^2|Au_{0}|^2)\\
&+& \frac{1}{\nu}\int_0^t e^{\frac{c}{\nu}\int_\ss^t(\|u^\aa_m(s)\|^2+\alpha^2|Au^\aa_m(s)|^2)ds}
|f(\ss)|^2\: d\ss,
\end{eqnarray*}
and by (\ref{tildeK0}) we obtain
\begin{equation}
\hskip-.0in
\|u^\aa_m(t)\|^2+\alpha^2 |A^\aa_mu(t)|^2 \le e^{\frac{c}{\nu^2}\tilde{\tilde{K}}_{01}^2}( \|u_0\|^2+
\alpha^2|Au_0|^2)+\frac{1}{\nu}e^{\frac{c}{\nu^2}\tilde{\tilde{K}}_{01}^2 }\int_0^T |f(\ss)|^2\: d\ss
=:\tilde{\tilde{K}}_{00}^2                                                                                                                            \label{tildeK01}.
\end{equation}
Integrate (\ref{pro-3}) and using (\ref{tildeK0}) and (\ref{tildeK01}), we obtan
\begin{eqnarray*}
\hskip-.2in
\|u^\aa_m(t)\|^2+\alpha^2 |Au^\aa_m(t)|^2
&+&\nu \int_0^t(|Au^\aa_m(s)|^2+\alpha^2|A^{3/2}u^\aa_m(s)|^2)\: ds  \\
&\le& \|u_0\|^2+\alpha^2|Au_0|^2+\frac{1}{\nu}\int_0^t |f|^2 \: dt +\frac{c}{\nu^2}\tilde{\tilde{K}}_{00}^2\tilde{\tilde{K}}_{01}^2,
\end{eqnarray*}
which concludes the proof.
\end{proof}
\begin{remark}
By following similar steps as those of NSE \cite{CF, Temam, Temam2}, one can show that the exact solutions of the 2D NS-$\alpha$ system have the same \textit{a priori} estimates as those of the solutions of the Galerkin system, namely, let $u^\aa$ be the solution of the system (\ref{NSa}) over the interval $[0,T]$ with initial data $u_0\in D(A)$, then
\begin{eqnarray}
\hskip-.8in
&& \underset{0\le t\le T}{\sup}\left[ |u^\aa(t)|^2+\alpha^2 \|u^\aa(t)\|^2 +\nu \int_0^t (\|u^\aa(s)\|^2
+\alpha^2 |Au^\aa(s)|^2) \: ds \right]\le \tilde{\tilde{K}}_{01}^2,            \label{NSa-estimates}  \\
\hskip-.8in
&&\underset{0\le t\le T}{\sup}\left[  \|u^\aa(t)\|^2+\alpha^2 |Au^\aa(t)|^2 +\nu \int_0^t (|A u^\aa(s)|^2+\alpha^2 |A^{3/2}u^\aa(s)|^2)\:ds        \right]   \le \tilde{\tilde{K}}_{02}^2.    \label{NSa-estimates2}
\end{eqnarray}
\end{remark}
\subsection{\textit{A priori} estimates for the 2D Modified Leray-$\alpha$ model}\label{subsecMLaapriori}
In this subsection, we will establish \textit{a priori} estimates for the 2D Modified Leray-$\alpha$ (ML-$\alpha$) regularization model of the 2D NSE.
Inspired by the NS-$\alpha$ and Leray-$\alpha$ models, the ML-$\alpha$ was introduced and analyzed in \cite{MLa} and is tested numerically in \cite{GKT} in the context of 3D sub-grid $\alpha$-models of turbulence.\\\\
The ML-$\alpha$ regularization model of the 2D NSE
is given by the
following functional evolution system in the space H:
\begin{eqnarray}
\hskip-.8in
\frac{dv^\aa}{dt}+\nu Av^\aa+B(v^\aa,u^\aa) &=& f,                        \label{MLa}   \\
u^\aa+\alpha^2 A u^\aa &=& v^\aa,      \nonumber \\
u^\aa(0)&=&u_0.                  \nonumber
\end{eqnarray}
Observe that when the regularization parameter $\alpha=0$ one again recovers the exact 2D NSE system (\ref{NSE}).\\\\
The corresponding Galerkin approximation system of the 2D ML-$\alpha$ model is given below as a system of ordinary differential equations in the space $H_m$:
\begin{eqnarray}
\hskip-.8in
\frac{dv^\aa_m}{dt}+\nu Av^\aa_m+P_mB(v^\aa_m,u^\aa_m) &=& f_m, \label{MLa-Galerkin}\\
u^\aa_m+\alpha^2 Au^\aa_m&=&v^\aa_m,  \nonumber  \\
u^\aa_m(0)&=&u_{0m}, \nonumber
\end{eqnarray}
where $u_{0m}=P_m\:u_0,\: f_m=P_m\:f$.\\\\
Following similar steps as in the $3D$ case in \cite{MLa} one can establish the global existence and uniqueness for the system (\ref{MLa}) in the case of $2D$ periodic boundary conditions.
In the next proposition we will present  \textit{a priori} estimates for the solutions of the Galerkin system
(\ref{MLa-Galerkin}).
\begin{proposition}\label{MLa-estimate}
Let $u_0\in D(A), T>0$ and $f \in L^2((0,T);H)$. Let $u^\aa_m$ be the solution of the Galerkin system of 2D ML-$\alpha$ (\ref{MLa-Galerkin}) over the interval $[0,T]$ with initial data $u_{0m}$ for a given $m\ge 1$, then
\begin{equation}
\hskip-.8in
\underset{0\le t\le T} {\sup} \left[ |u^\aa_m(t)|^2+\alpha^2 \|u^\aa_m(t)\|^2 +\nu \int_0^t (\|u^\aa_m(s)\|^2+\alpha^2|Au^\aa_m(s)|^2) \:ds
\right]
\le \tilde{\tilde{K}}_{01}^2.     \label{tildeK-0}
\end{equation}
Moreover,
\begin{equation}
\hskip-.8in
\underset{0\le t\le T} {\sup}\left[\|u^\aa_m(t)\|^2+\alpha^2 |Au^\aa_m(t)|^2 +\nu \int_0^t (|Au^\aa_m(s)|^2+\alpha^2 |A^{3/2}u^\aa_m(s)|^2) \:ds \right]\le \tilde{\tilde{K}}_{02}^2,   \label{ttildeK-0}
\end{equation}
where $\tilde{\tilde{K}}_{01}$ and $\tilde{\tilde{K}}_{02}$ are the same as in Proposition \ref{NSa-estimate}.
\end{proposition}
\begin{proof}
Let us take the inner product of the equation in (\ref{MLa-Galerkin}) with $u^\aa_m$ to obtain
\begin{eqnarray*}
\hskip-.2in
\frac{1}{2}\frac{d}{dt}(|u^\aa_m|^2+\alpha^2 \|u^\aa_m\|^2) +\nu( \|u^\aa_m\|^2+\alpha^2 |Au^\aa_m|^2)
&=& (P_mf,u^\aa_m) \le |f||u^\aa_m|
\le|f| \frac{\|u^\aa_m\|}{ \lambda_1^{1/2}}  \\
&\le& \frac{|f|^2}{2\nu \lambda_1}+\frac{\nu}{2}\|u^\aa_m\|^2,
\end{eqnarray*}
thus we get
\begin{equation}
\hskip-.8in
\frac{d}{dt}(|u^\aa_m|^2+\alpha^2 \|u^\aa_m\|^2) +\nu( \|u^\aa_m\|^2+\alpha^2|Au^\aa_m|^2) \le \frac{|f|^2}{\nu \lambda_1}.  \label{MLa-error1}
\end{equation}
Integrating the above over the interval $(0,t)$, we obtain
\begin{equation*}
\hskip-.0in
|u^\aa_m(t)|^2+\alpha^2 \|u^\aa_m(t)\|^2+\nu \int_0^t (\|u^\aa_m(s)\|^2+\alpha^2 |Au^\aa_m(s)|^2 ) \: ds \: \le \int_0^t \frac{|f(s)|^2}{\nu\lambda_1} \: ds+|u_0|^2+\alpha^2\|u_0\|^2,
\end{equation*}
which implies (\ref{tildeK-0}).\\\\
Next, we take the inner product of the equation in (\ref{MLa-Galerkin}) with $Au^\aa_m$ to obtain
\begin{eqnarray}
\hskip-.3in
\frac{1}{2}\frac{d}{dt}(\|u^\aa_m\|^2+\alpha^2|Au^\aa_m|^2)
& +&\nu( |Au^\aa_m|^2+\alpha^2 |A^{3/2}u^\aa_m|^2)  \nonumber \\
&=&   -(B(v^\aa_m,u^\aa_m),Au^\aa_m)+(P_mf,Au^\aa_m) \nonumber \\
&\le& |(B(v^\aa_m,u^\aa_m),Au^\aa_m)|+\frac{|f|^2}{2\nu}+\frac{\nu}{2}|Au^\aa_m|^2.   \label{MLa-error2}
\end{eqnarray}
Next we estimate
\begin{equation*}
\hskip-.8in
|B(v^\aa_m,u^\aa_m),Au^\aa_m)|=|(B(u^\aa_m,u^\aa_m),Au^\aa_m)+\alpha^2(B(Au^\aa_m,u^\aa_m),Au^\aa_m)|.
\end{equation*}
By (\ref{b-uuau}) and Proposition 7 we have
\begin{eqnarray}
\hskip-.8in
|(B(v^\aa_m,u^\aa_m),Au^\aa_m)|
&=& \alpha^2|(B(Au^\aa_m,u^\aa_m),Au^\aa_m)| \le c\alpha^2 |A^{3/2}u^\aa_m| \|u^\aa_m\| |Au^\aa_m|   \nonumber \\
&\le& \frac{\nu}{2}\alpha^2 |A^{3/2}u^\aa_m|^2+\frac{c}{2\nu} \|u^\aa_m\|^2(\alpha^2 |Au^\aa_m|^2) \nonumber \\
&\le& \frac{\nu}{2} \alpha^2 |A^{3/2}u^\aa_m|^2+\frac{c}{2\nu} (\|u^\aa_m\|^2+\alpha^2 |Au^\aa_m|^2 )^2.  \label{MLa-error3}
\end{eqnarray}
From above we have
\begin{equation*}
\hskip-.2in
\frac{d}{dt}(\|u^\aa_m\|^2+\alpha^2 |Au^\aa_m|^2) +\nu (|Au^\aa_m|^2+\alpha^2 |A^{3/2} u^\aa_m|^2) \le \frac{|f|^2}{\nu}+\frac{c}{\nu}
(\|u^\aa_m\|^2+\alpha^2 |Au^\aa_m|^2)^2,
\end{equation*}
which is exactly the same inequality as (\ref{pro-3}) (as in the proof of Proposition \ref{NSa-estimate}).
Following the same steps we obtain (\ref{ttildeK-0}).
\end{proof}
\begin{remark}
Similarly, the solutions of the exact ML-$\alpha$ system have the same \textit{a priori} estimates as those of the solutions of the Galerkin system, namely, let $u^\aa$ be the solution of the system (\ref{MLa}) over the interval $[0,T]$ with initial data $u_0\in D(A)$, then
\begin{equation}
\hskip-.8in
\underset{0\le t\le T} {\sup} \left[ |u^\aa(t)|^2+\alpha^2 \|u^\aa(t)\|^2 +\nu \int_0^t (\|u^\aa(s)\|^2+\alpha^2|Au^\aa(s)|^2) \:ds
\right]
\le \tilde{\tilde{K}}_{01}^2.     \label{MLa-apriori}
\end{equation}
Moreover,
\begin{equation}
\hskip-.8in
\underset{0\le t\le T} {\sup}\left[\|u^\aa(t)\|^2+\alpha^2 |Au^\aa(t)|^2 +\nu \int_0^t (|Au^\aa(s)|^2+\alpha^2 |A^{3/2}u^\aa(s)|^2) \:ds \right]\le \tilde{\tilde{K}}_{02}^2,   \label{MLa-apriori2}
\end{equation}
where $\tilde{\tilde{K}}_{01}$ and $\tilde{\tilde{K}}_{02}$ are the same as in Proposition \ref{NSa-estimate}.
\end{remark}
\subsection{\textit{A priori} estimates for the 2D simplified Bardina model}\label{subsecBardinaapriori}
In this subsection, we will establish \textit{a priori} estimates for the 2D simplified Bardina model for
the 2D NSE.
The
Bardina closure model of turbulence was introduced first by Bardina et al \cite{OBardina} and later simplified
and studied further in \cite{Layton} and in \cite{Bardina}.
In particular, global well-posedness for the $3D$ simplified Bardina model was established in \cite{Layton}. In \cite{Bardina} these results were slightly improved and the long-time behavior was investigated. Further, it was established in \cite{Bardina} the global regularity of the $3D$ inviscid version of the simplified Bardina, a property which is still out of reach for the other $\alpha$-models of
turbulence.\\\\
The simplified Bardina regularization model of 2D NSE is given by the following functional
evolution system in the space H:
\begin{eqnarray}
\hskip-.8in
\frac{dv^\aa}{dt}+\nu Av^\aa+B(u^\aa,u^\aa)&=&f,           \label{Bardina}   \\
u^\aa+\alpha^2Au^\aa &=& v^\aa,  \nonumber \\
u^\aa(0)&=&u_0.              \nonumber
\end{eqnarray}
Observe again that when the regularization parameter $\alpha=0$ one recovers the exact 2D NSE system (\ref{NSE}).\\\\
The corresponding Galerkin approximation system of the 2D simplified Bardina model is given below as a system of ordinary differential equations in
space $H_m$:
\begin{eqnarray}
\hskip-.8in
\frac{dv^\aa_m}{dt}+\nu Av^\aa_m+P_mB(u^\aa_m,u^\aa_m) &=& f_m, \label{Bardina-Galerkin} \\
u^\aa_m+\alpha^2 Au^\aa_m &=& v^\aa_m, \nonumber \\
u^\aa_m(0)&=& u_{0m}, \nonumber
\end{eqnarray}
where $u_{0m}=P_m\: u_0,\: f_m=P_m\:f$.\\\\
In the following proposition we will establish \textit{a prior} estimates for the solutions of the Galerkin system (\ref{Bardina-Galerkin}).
\begin{proposition}\label{Bardina-estimate}
Let $u_0 \in D(A), T>0$ and $f \in L^2((0,T);H)$. Let $u^\aa_m$ be the solution of the system (\ref{Bardina-Galerkin}) over the interval $[0,T]$
with initial data $u_{0m}$ for a given $m\ge 1$, then
\begin{equation*}
\hskip-.2in
\underset{0\le t \le T}{\sup} \left[ \|u^\aa_m(t)\|^2+\alpha^2 |Au^\aa_m(t)|^2+\nu \int_0^t( |Au^\aa_m(s)|^2+\alpha^2|A^{3/2}u^\aa_m(s)|^2 )
\: ds \right] \le \tilde{K}_0^2,
\end{equation*}
where $\tilde{K}_0$ is given in Proposition \ref{LA-estimate}.
\end{proposition}
\begin{proof}
We take the inner product of the first equation in system (\ref{Bardina-Galerkin}) with $Au^\aa_m$ and using (\ref{b-uuau}) to obtain
\begin{equation*}
\hskip-.2in
\frac{1}{2}\frac{d}{dt}(\|u^\aa_m\|^2+\alpha^2 |Au^\aa_m|^2)+\nu( |Au^\aa_m|^2+\alpha^2 |A^{3/2}u^\aa_m|^2)=(P_mf,Au^\aa_m)
\le \frac{\nu}{2} |Au^\aa_m|^2+\frac{|f|^2}{2\nu}.
\end{equation*}
Hence
\begin{equation*}
\frac{d}{dt} (\|u^\aa_m\|^2+\alpha^2|Au^\aa_m|^2 )+\nu (|Au^\aa_m|^2+\alpha^2 |A^{3/2} u^\aa_m|^2) \le \frac{|f|^2}{\nu}.
\end{equation*}
Integrating the above from $0$ to $t$, we have
\begin{equation*}
\hskip-.1in
\|u^\aa_m(t)\|^2+\alpha^2|Au^\aa_m(t)|^2+\nu\int_0^t ( |Au^\aa_m(s)|^2+\alpha^2 |A^{3/2}u^\aa_m(s)|^2) \: ds
\le \frac{1}{\nu}\int_0^T |f(s)|^2\: ds +\|u_0\|^2+\alpha^2 |Au_0|^2,
\end{equation*}
which concludes our proof.
\end{proof}
\begin{remark}
Similar to the cases of the other $\alpha$-regularization models, one can show that the solutions of the simplified Bardina model have the same \textit{a priori} estimates as those of the Galerking system, namely, let $u^\aa$ be the solution of the system (\ref{Bardina}) over the interval $[0,T]$ with initial data $u_0\in D(A)$, then
\begin{equation}
\hskip-.8in
\underset{0\le t \le T}{\sup} \left[ \|u^\aa(t)\|^2+\alpha^2 |Au^\aa(t)|^2+\nu \int_0^t( |Au^\aa(s)|^2+\alpha^2|A^{3/2}u^\aa(s)|^2 )
\: ds \right] \le\tilde{K}_0^2,  \label{Bardina-apriori}
\end{equation}
where $\tilde{K}_0$ is given in Proposition \ref{LA-estimate}.\\
\end{remark}
\section{Rates of Convergence of $\alpha$-Regularization Models to the Navier-Stokes Equations}\label{errorconv}
We aim here to show the convergence rates of solutions of the
various $\alpha$-models toward the corresponding solution of the
exact NSE equation when the regularization parameter $\alpha$
approaches zero. We will focus on the $L^2$-norm of the difference
between $u^\aa$, solution of the underlying $\alpha$-model,  and
$u$, solution of the exact NSE. From the results concerning all four
$\alpha$-models included in this study, we observe that all the
errors $|u-u^\aa|$ are of the same order of
$O(\frac{\alpha}{L}(\log\frac{L}{\alpha})^{1/2})$, as
$\frac{\alpha}{L}$ goes to zero. Though the four $\alpha$-models we
investigate in this paper share the same order of error estimates
with the only difference in the constant, the treatment of each
nonlinearity is slightly different and we simply present the details
for the readers' benefit. The similar error estimates of these four
$\alpha$-models show that, essentially, these four regularization
models are consistent in their convergence to the exact 2D NSE
system.
\\\\
Before we proceed in showing the main results of convergence rates of the $\alpha$-models, we state an important lemma and a proposition which will play a crucial role in estimating the convergence rates of the various
approximation models.
\begin{lemma}\label{general-error}
Let $\alpha>0$ be a given fixed parameter, and
let $\vp \in H$ and $\dd \in V$. Then
\begin{equation*}
\hskip-.8in
| ( \vp -(I+\alpha^2 A)^{-1} \vp, \dd )| \le \frac{\alpha}{2} | \vp | \| \dd \|.
\end{equation*}
\end{lemma}
\begin{proof}
First we observe that
\begin{equation*}
\hskip-.8in
\vp-(I+\alpha^2 A)^{-1} \vp
=\alpha^2 A(I+\alpha^2A)^{-1} \vp.
\end{equation*}
Therefore,
\begin{eqnarray*}
\hskip-.8in
| (\vp-(I+\alpha^2A)^{-1}\vp, \dd)|
&&=|(\alpha^2A(I+\alpha^2A)^{-1}\vp, \dd)|                                                                                         \\
&&=|\alpha \left ( (\alpha^2 A)^{1/2}(I+\alpha^2A)^{-1} \vp, A^{1/2} \dd \right) |                       \\
&& \le \alpha | (\alpha^2 A)^{1/2} (I+\alpha^2A)^{-1} \vp | \| \dd\|                                                \\
&&\le \alpha \| (\alpha^2 A)^{1/2}(I+\alpha^2 A)^{-1} \| _{\mathcal{L}(H)} | \vp | \| \dd \|           \\
&&\le \frac{\alpha}{2} | \vp | \| \dd\|,
\end{eqnarray*}
where the last inequality follows from the fact that the operator norm
\begin{eqnarray*}
\hskip-.8in
\| (\alpha^2A)^{1/2} (I+\alpha^2A)^{-1} \|_{    \mathcal{L}(H)}
&&=\underset{k=1,2,...}{\sup} \left(   \frac{ (\alpha^2\lambda_k)^{1/2}}{ 1+\alpha^2 \lambda_k    }              \right)                                 \\
&&\le \underset{0\le y< \infty} {\sup} \frac{ y}{1+y^2} =\frac{1}{2},
\end{eqnarray*}
which concludes our proof.
\end{proof}
\begin{proposition}\label{infinity-inequality}
Let $u_0\in D(A)$ and $T>0$. Assume that $\alpha$ is small enough such that $\frac{L}{2\pi \alpha} \ge 1$, and suppose that $u^\aa(t)$ is the solution of system (\ref{LA}) or (\ref{Bardina}) , then
\begin{equation}
\hskip-.8in
\underset{0\le t\le T}{\sup} \|u^\aa(t)\|_{L^{\infty}}^2 \le c \tilde{K}_0^2\left(1+\log\left(\frac{L}{2\pi \alpha}\right) \right).
\end{equation}
\end{proposition}
\begin{proof}
By the Brezis-Gallouet inequality in Proposition \ref{BG-inequality}, we have
\begin{eqnarray*}
\hskip-.2in
\|u^\aa(t)\|_{L^{\infty}}^2
&\le& c\|u^\aa(t)\|^2 \left( 1+\log \left( \frac{L}{2\pi} \frac{|Au^\aa(t)|}{\|u^\aa(t)\|}  \right) \right)      \\
&\le&  c\left(   \|u^\aa(t)\|^2+\|u^\aa(t)\|^2\log\left(\frac{L}{2\pi \alpha}\right)+\|u^\aa(t)\|^2 \log\left(  \frac{\alpha|Au^\aa(t)|}{\|u^\aa(t)\|}     \right)              \right).
\end{eqnarray*}
Therefore, by Proposition \ref{LA-estimate} for system (\ref{LA}) or Proposition \ref{Bardina-estimate} for system (\ref{Bardina}), we have $\alpha |Au^\aa(t)| \le \tilde{K}_0$ and $\|u^\aa(t)\| \le \tilde{K}_0$, for all $t\in[0,T]$. Since $\frac{L}{2\pi\alpha} \ge 1$, we obtain
\begin{equation*}
\hskip-.8in
\|u^\aa(t)\|_{L^{\infty}}^2 \le c \left(  \tilde{K}_0^2+\tilde{K}_0^2\log \left( \frac{L}{2\pi \alpha} \right)+\tilde{K}_0^2 \left(\frac{\|u^\aa(t)\|}{\tilde{K}_0} \right)
\log \left(\frac{\tilde{K}_0}{\|u^\aa(t)\|}\right)                                         \right).
\end{equation*}
Now, since by Proposition \ref{LA-estimate} or Proposition \ref{Bardina-estimate},   \: $\|u^\aa(t)\|\le \tilde{K}_0 $, for all $t\in [0,T]$, then we have
\begin{equation*}
\hskip-.8in
\left(\frac{\|u^\aa(t)\|}{\tilde{K}_0}\right) \log \left(\frac{\tilde{K}_0}{\|u^\aa(t)\|}\right) \le \frac{1}{e}, \qquad  \mbox{for all } t\in [0,T].
\end{equation*}
Consequently, from all the above we have
\begin{equation*}
\hskip-.8in
\underset{0\le t\le T}{\sup} \|u^\aa(t)\|_{L^{\infty}}^2
\le c \tilde{K}_0^2 \left(  1+\log \left(\frac{L}{2\pi\alpha}\right)     \right).
\end{equation*}
\end{proof}
\begin{remark}
The result above also holds for the solutions $u^\aa$ of systems (\ref{NSa}) and (\ref{MLa}) with $\tilde{K}_0$ replaced by $\tilde{\tilde{K}}_{02}$ in the cases of the NS-$\alpha$ and the
ML-$\alpha$ models.
\end{remark}
Now we are ready to present the convergence rates of various $\alpha$ turbulence models.
\subsection{The rate of convergence of the Leray-$\alpha$ model}\label{subsecLAconv}
In this subsection, we will establish error estimates concerning the rate of convergence of the solutions
of the 2D Leray-$\alpha$ model to solutions of the 2D NSE, as the regularization parameter $\alpha$ goes to zero. We will proceed by estimating the $L^2$-norm of the difference $\delta^\aa=u-u^\aa$, where $u$ is the solution of 2D NSE system (\ref{NSE}) and $u^\aa$ is the solution of 2D Leray-$\alpha$ model (\ref{LA}).\\\\
From (\ref{NSE}) and (\ref{LA}) we observe that $\delta^\aa=u-u^\aa$ satisfies the following equation:
\begin{equation}
\frac{d}{dt}\dd^\aa+\nu A\dd^\aa +B(u,\dd^\aa)+B(\dd^\aa, u)-B(\dd^\aa,\dd^\aa) -(I+\alpha^2 A)^{-1}B(u^\aa,v^\aa)+B(u^\aa,u^\aa)=f-(I+\alpha^2A)^{-1}f.      \label{LA-delta1}
\end{equation}
\begin{theorem} \label{LA-theorem}
Let $u^\aa$ be a solution of the 2D Leray-$\alpha$ system (\ref{LA}) with initial data $u_0\in D(A)$, and let
$u$ be the solution of the 2D NSE system (\ref{NSE}) with the same initial data $u_0$ over the interval $[0,T]$. Assume that $\alpha$ is small
enough such that $\frac{L}{2\pi \alpha}\ge 1$ and let $\delta^\aa=u-u^\aa$, then
\begin{equation*}
\hskip-.8in
\underset{0\le t\le T} {\sup}|\dd^\aa(t)|^2 \le \epsilon ^2,
\end{equation*}
where
$\epsilon^2:=\frac{c\alpha^2}{\nu}e^{\frac{cK_0^2}{\nu^2}} \left(  T\tilde{K}_0^4\left( 1+\log\left( \frac{L}{2\pi \alpha} \right  )         \right) + \|f\|_{L^2((0,T);H)}^2    \right)$, $\tilde{K}_0$ and $K_0$ are given in Proposition
\ref{LA-estimate} and \ref{NSE-estimate} respectively, and here $c$ is a dimensionless constant that is independent of $\nu$, $\alpha$, $f$ or $T$.
\end{theorem}
\begin{proof}
We take the inner product of the above equation (\ref{LA-delta1})  with $\dd^\aa$ and use the identity (\ref{b-uvv}) to obtain
\begin{eqnarray*}
\hskip-.2in
\frac{1}{2}\frac{d}{dt}|\dd^\aa|^2
&+&\nu \|\dd^\aa \|^2+(B(\dd^\aa, u), \dd^\aa)+(   (I+\alpha^2A)^{-1}(B(u^\aa,u^\aa)-B(u^\aa,v^\aa)),  \dd^\aa )       \\
&+&(    ( I - (I+\alpha^2A)^{-1})B(u^\aa,u^\aa), \dd^\aa )=(  (I-(I+\alpha^2A)^{-1})f,\dd^\aa    ).
\end{eqnarray*}
Consequently, one has
\begin{equation}
\hskip-.8in
\frac{1}{2}\frac{d}{dt}|\dd^\aa|^2 +\nu \|\dd^\aa \|^2 \le |J_1|+|J_2|+|J_3|+|J_4|,  \label{Js}
\end{equation}
where
\begin{eqnarray}
&&\hskip-.8in
J_1=(B(\dd^\aa, u), \dd^\aa),          \label{J-1}       \\
&&\hskip-.8in
J_2=\left(  (I+\alpha^2A)^{-1}(B(u^\aa,u^\aa)-B(u^\aa,v^\aa) ), \dd^\aa   \right),    \label{J-2}    \\
&&\hskip-.8in
J_3=\left(  (I-(I+\alpha^2A)^{-1} )B(u^\aa,u^\aa),\dd^\aa \right),  \label{J-3}               \\
&&\hskip-.8in
J_4=\left(  (I-(I+\alpha^2A)^{-1})f, \dd^\aa          \right).      \label{J-4}
\end{eqnarray}
Next, we estimate $|J_1|,\:|J_2|,\: |J_3|$ and $|J_4|$. Thanks to estimate (\ref{b-uvw}) and Young's inequality we have
\begin{equation}
\hskip-.8in
|J_1|=|(B(\dd^\aa, u), \dd^\aa)| \le c|\dd^\aa| \| \dd^\aa \| \| u\| \le \frac{\nu}{8}\| \dd^\aa \|^2+\frac{c}{\nu}|\dd^\aa|^2 \|u\|^2.
\label{J-1-estimate}
\end{equation}
Since $u^\aa+\alpha^2 A u^\aa=v^\aa$, we obtain from (\ref{J-2})
\begin{eqnarray*}
\hskip-.8in
|J_2|
&=&|((I+\alpha^2A)^{-1}B(u^\aa,  \alpha^2Au^\aa), \dd^\aa)|  \\
&=& |(B(u^\aa, \alpha^2 Au^\aa), (I+\alpha^2A)^{-1}\dd^\aa)|,
\end{eqnarray*}
and by applying (\ref{bilinear}), (\ref{b-infty}) and Young's inequality we have
\begin{eqnarray*}
\hskip-.8in
|J_2|
&=&|(B(u^\aa, (I+\alpha^2 A)^{-1}\dd^\aa), \alpha^2Au^\aa)|       \\
&\le&c\alpha^2 \|u^\aa\|_{L^{\infty}} \|(I+\alpha^2A)^{-1}\dd^\aa \| |Au^\aa|   \\
&\le&c\alpha^2 \|u^\aa\|_{L^{\infty}} |Au^\aa| \|\dd^\aa\|          \\
&\le&\frac{\nu}{8}\| \dd^\aa\|^2 +\frac{c\alpha^4}{\nu} |Au^\aa|^2 \|u^\aa\|_{L^{\infty}}^2.
\end{eqnarray*}
From Proposition \ref{LA-estimate}, we have $\underset{0\le t\le T}{\sup}\alpha^2 |Au^\aa|^2 \le \tilde{K}_0^2$, then by further applying Proposition
\ref{infinity-inequality} we have
\begin{equation}
\hskip-.8in
|J_2|\le \frac{\nu}{8}\| \dd^\aa\|^2 +\frac{c\alpha^2}{\nu} \tilde{K}_0^4\left(1+\log \left( \frac{L}{2\pi \alpha}\right)\right ).    \label{J-2-estimate}
\end{equation}
Now, we are ready to estimate $|J_3|$.\\
By Lemma \ref{general-error}, (\ref{J-3}) and Young's inequality we have
\begin{eqnarray*}
\hskip-.8in
|J_3|
&\le&\frac{\alpha}{2}|B(u^\aa,u^\aa)| \| \dd^\aa\| \le \frac{\nu}{8} \| \dd^\aa\|^2+\frac{\alpha^2}{2\nu}|B(u^\aa,u^\aa)|^2   \\
&\le&\frac{\nu}{8}\|\dd^\aa\|^2 +\frac{c\alpha^2}{\nu}\|u^\aa\|_{L^{\infty}}^2 \|u^\aa\|^2.
\end{eqnarray*}
Since $\underset{0\le t\le T}{\sup}\|u^\aa(t)\|^2\le \tilde{K}_0^2$ by Proposition \ref{LA-estimate}, then by Proposition \ref{infinity-inequality} we reach
\begin{equation}
\hskip-.8in
|J_3|\le\frac{\nu}{8}\|\dd^\aa\|^2+\frac{c\alpha^2}{\nu}\tilde{K}_0^4 \left(1+\log \left(\frac{L}{2\pi\alpha} \right)\right).  \label{J-3-estimate}
\end{equation}
Finally, we estimate $|J_4|$. By Lemma \ref{general-error}, (\ref{J-4}) and Young's inequality we obtain
\begin{equation}
\hskip-.8in
|J_4|\le \frac{\alpha}{2}|f| \| \dd^\aa \| \le \frac{\nu}{8}\| \dd^\aa \|^2+\frac{\alpha^2}{2\nu}|f|^2.   \label{J-4-estimate}
\end{equation}
Now from (\ref{Js}), (\ref{J-1-estimate}), (\ref{J-2-estimate}), (\ref{J-3-estimate}) and (\ref{J-4-estimate})
we have
\begin{equation*}
\hskip-.8in
\frac{d}{dt}|\dd^\aa|^2+\nu\| \dd^\aa\|^2 \le \frac{c}{\nu}\|u\|^2|\dd^\aa|^2+\frac{c\alpha^2}{\nu}
\left(|f|^2+\tilde{K}_0^4\left(1+\log \left(\frac{L}{2\pi \alpha}\right)\right)\right).
\end{equation*}
Dropping the $\nu\|\dd^\aa\|^2$ term from the left-hand side and applying Gronwall's inequality we obtain
\begin{equation*}
\hskip-.1in
|\dd^\aa(t)|^2 \le e^{\frac{c}{\nu}\int_0^t\|u(\sigma)\|^2d\sigma } |\dd^\aa(0)|^2 +\frac{c\alpha^2}{\nu}
\int_0^t e^{\frac{c}{\nu}\int_{\sigma}^t\|u(s)\|^2ds} \left[|f(\sigma)|^2+\tilde{K}_0^4\left(1+\log\left(\frac{L}{2\pi\alpha}\right) \right)\right]\: d\sigma.
\end{equation*}
Since $\dd^\aa(0)=0$, then by Propositon \ref{NSE-estimate} we obtain
\begin{equation*}
\hskip-.8in
|\dd^\aa(t)|^2 \le \frac{c\alpha^2}{\nu}e^{\frac{cK_0^2}{\nu^2}}
\left(    T\tilde{K}_0^4\left(1+\log\left(\frac{L}{2\pi\alpha}\right)\right)
+\|f\|_{L^2((0,T);H)}^2             \right)=:\epsilon^2,
\end{equation*}
which concludes our proof.
\end{proof}
\subsection{The rate of convergence of the Navier-Stokes-$\alpha$ model}\label{subsecNSaconv}
In this subsection, we will establish error estimates concerning the rate of convergence of solutions
of the 2D NS-$\alpha$ regularization model to solutions of the 2D NSE, as the regularization parameter $\alpha$ goes to zero. We will proceed by estimating the $L^2$-norm of the difference $\dd^\aa=u-u^\aa$, where $u$ is the solution of 2D NSE system (\ref{NSE}) and $u^\aa$ is the solution of 2D NS-$\alpha$
model (\ref{NSa}).\\\\
From (\ref{NSE}) and (\ref{NSa}) we observe that $\delta^\aa=u-u^\aa$ satisfies the following equation:
\begin{eqnarray}
\hskip-.4in
&&\frac{d}{dt}\dd^\aa
+\nu A\dd^\aa
+B(\dd^\aa,u)+B(u,\dd^\aa)-B(\dd^\aa,\dd^\aa)
+(I+\alpha^2A)^{-1}(B(u^\aa,u^\aa)-B(u^\aa,v^\aa)) \nonumber \\
&&+(I-(I+\alpha^2A)^{-1})B(u^\aa,u^\aa)+(I+\alpha^2A)^{-1}B^*(u^\aa,v^\aa)=(I-(I+\alpha^2A)^{-1})f,    \label{NSa-erroreqn1}
\end{eqnarray}
where for every $v\in V$ fixed, the operator $B^*(\cdot, v)$ is the adjoint operator of the operator
$B(\cdot, v)$, which is defined by $(B^*(\psi,v),\vp)=(B(\vp,v),\psi)$ for every $\vp, \psi \in \mathcal{V}$ (see relation (\ref{tildeB-id})).
\begin{theorem}\label{NSa-error-theorem}
Let $u^\aa$ be the solution of the 2D NS-$\alpha$ system (\ref{NSa}) with initial data $u_0\in D(A)$, and let $u$ be the solution of the 2D NSE system
(\ref{NSE}) with the same initial data $u_0$ over the interval $[0,T]$.  Assume that $\alpha$ is small enough such that
$\frac{L}{2\pi \alpha}\ge 1$ and let $\dd^\aa=u-u^\aa$, then
\begin{equation*}
\hskip-.8in
\underset{0\le t\le T}{\sup}|\dd^
\aa(t)|^2 \le \tilde{\epsilon}^2,
\end{equation*}
where $\tilde{\epsilon}^2:=\frac{c\alpha^2 }{\nu} e^{\frac{cK_0^2}{\nu^2}}
\left( T\tilde{\tilde{K}}_{02}^4 \left(1+\log \left(\frac{L}{2\pi\alpha}\right) \right) +\|f\|^2_{L^2((0,T);H)} \right)$, and $K_0$, $\tilde{\tilde{K}}_{02} $ are given in (\ref{K-1}) and (\ref{ttildeK0}), respectively.
\end{theorem}
\begin{proof}
By taking inner product of (\ref{NSa-erroreqn1}) with $\dd^\aa$ and
using (\ref{b-uvv}) and (\ref{b-uuau}) we have
\begin{equation*}
\hskip-.8in
\frac{1}{2}\frac{d}{dt} |\dd^\aa|^2+ \nu \|\dd^\aa\|^2 +J_1+J_2+J_3+J_5=J_4,
\end{equation*}
where $J_1,J_2,J_3$ and $J_4$ are given in (\ref{J-1})-(\ref{J-4}) and $J_5=((I+\alpha^2A)^{-1}B^*(u^\aa,v^\aa),\dd^\aa)$.
Therefore,
\begin{equation}
\hskip-.8in
\frac{1}{2}\frac{d}{dt}|\dd^\aa|^2+\nu \|\dd^\aa\|^2 \le |J_1|+|J_2|+|J_3|+|J_4|+|J_5|.   \label{NSa-error1}
\end{equation}
The estimates for $|J_1|, |J_2|,|J_3|$ and $|J_4|$ follow exactly the same steps as in the proof of
Theorem \ref{LA-theorem}, where $\tilde{K}_0$ is replaced by $\tilde{\tilde{K}}_{02}$ from Proposition \ref{NSa-estimate} (see also Remark 6).\\\\
Next, we estimate $|J_5|$. To this end we first observe that $B^*(u,u) \equiv 0$, because, thanks to (\ref{b-uvv}),
\begin{equation*}
\hskip-.8in
\langle B^*(u,u),w\rangle_{V'}=\langle B(w,u),u\rangle_{V'}=0  \quad \quad \mbox{for every } w\in V.
\end{equation*}
Thus,
\begin{eqnarray*}
\hskip-.8in
|J_5|
&=& |((I+\alpha^2A)^{-1}B^*(u^\aa,v^\aa),\dd^\aa)| \\
&=&|((I+\alpha^2A)^{-1}(B^*(u^\aa,v^\aa)-B^*(u^\aa,u^\aa)),\dd^\aa)| \\
&=&\alpha^2 |((I+\alpha^2A)^{-1}B^*(u^\aa,Au^\aa),\dd^\aa)|,
\end{eqnarray*}
where we used the relation $u^\aa+\alpha^2Au^\aa=v^\aa$ and the bilinearity of $B^*(\cdot, \cdot)$.\\
Consequently, by the definition of $B^*$ and (\ref{bilinear}), we have
\begin{equation*}
\hskip-.8in
|J_5|=\alpha^2|(B((I+\alpha^2A)^{-1}\dd^\aa, Au^\aa),u^\aa)|=\alpha^2|(B((I+\alpha^2A)^{-1}\dd^\aa,u^\aa),Au^\aa)|.
\end{equation*}
Therefore, by Lemma \ref{B-lemma} and Young's inequality we have
\begin{eqnarray*}
\hskip-.8in
|J_5| &\le& c\alpha^2 \|(I+\alpha^2A)^{-1} \dd^\aa\| \|u^\aa\| |Au^\aa|  \\
&\le&c\alpha^2 \| \dd^\aa\| \|u^\aa\| |Au^\aa| \\
&\le& \frac{\nu \|\dd^\aa\|^2}{8}+\frac{c\alpha^4}{\nu}\|u^\aa\|^2|Au^\aa|^2.
\end{eqnarray*}
By the above and Proposition \ref{NSa-estimate} we conclude
\begin{equation}
\hskip-.8in
|J_5|\le \frac{\nu\|\dd^\aa\|^2}{8}+\frac{c\alpha^2}{\nu}\tilde{\tilde{K}}_{02}^4.  \label{J-5-estimate}
\end{equation}
From (\ref{J-1-estimate})-(\ref{J-4-estimate}) (where $\tilde{K}_0$ is replaced by $\tilde{\tilde{K}}_{02}$, see also Remark 6),
(\ref{NSa-error1}) and (\ref{J-5-estimate}) we obtain
\begin{equation*}
\hskip-.8in
\frac{d}{dt}|\dd^\aa|^2 +\nu \|\dd^\aa\|^2 \le \frac{c}{\nu}\|u\|^2 |\dd^\aa|^2+\frac{c}{\nu}\alpha^2
\left(|f|^2+\tilde{\tilde{K}}_{02}^4 \left(1+\log \left( \frac{L}{2\pi\alpha} \right) \right)\right).
\end{equation*}
Dropping the $\nu \|\dd^\aa\|^2$ term from the left-hand side and applying Gronwall's inequality we obtain
\begin{equation*}
\hskip-.0in
|\dd^\aa(t)|^2 \le e^{\frac{c}{\nu}\int_0^t \|u(\ss)\|^2 d\ss} |\dd^\aa(0)|^2+\frac{c\alpha^2}{\nu}
\int_0^t e^{\frac{c}{\nu} \int_\ss^t \|u(s)\|^2ds} \left(|f(\ss)|^2+
\tilde{\tilde{K}}_{02}^4\left(1+\log\left( \frac{L}{2\pi\alpha}\right) \right) \right)\: d\ss.
\end{equation*}
Since $\dd^\aa(0)=0$, then by Proposition \ref{NSE-estimate} and the above we obtain
\begin{equation*}
\hskip-.8in
|\dd^\aa(t)|^2 \le \frac{c\alpha^2}{\nu} e^{ \frac{cK_0^2}{\nu^2}} \left( T\tilde{\tilde{K}}_{02}^4
\left(1+\log \left( \frac{L}{2\pi \alpha}\right)\right) +\|f\|_{L^2((0,T);H)}^2 \right) =:\tilde{\epsilon}^2.
\end{equation*}
\end{proof}
\subsection{The rate of convergence of the Modified Leray-$\alpha$ model}\label{subsecMLaconv}
In this subsection, we will establish error estimates concerning the rate of convergence of solutions of the 2D ML-$\alpha$ regularization model to solutions of the 2D NSE, as the regularization parameter $\alpha$ goes to zero. We will proceed by estimating the $L^2$-norm of the difference $\dd^\aa=u-u^\aa$, where $u$ is the solution of 2D NSE system (\ref{NSE}) and $u^\aa$ is the solution of the 2D ML-$\alpha$ model (\ref{MLa}).\\\\
From (\ref{NSE}) and (\ref{MLa}) we observe that $\delta^
\aa=u-u^\aa$ satisfies the following equation:
\begin{eqnarray}
\hskip-.8in
\frac{d}{dt}\dd^\aa
&+& B(\dd^\aa,u)+B(u,\dd^\aa)-B(\dd^\aa,\dd^\aa)+(I+\alpha^2A)^{-1}(B(u^\aa,u^\aa)-B(v^\aa,u^\aa))  \nonumber \\
&+&(I-(I+\alpha^2A)^{-1})B(u^\aa,u^\aa)=(I-(I+\alpha^2A)^{-1}) f.   \label{MLa-delta1}
\end{eqnarray}
\begin{theorem}\label{MLa-error}
Let $u^\aa$ be the solution of the 2D Modified Leray-$\alpha$ system (\ref{MLa}) with initial data $u_0\in D(A)$, and let
$u$ be the solution of the 2D NSE system (\ref{NSE}) with the same initial data $u_0$ over the interval $[0,T]$. Assume that $\alpha$ is small enough such that $\frac{L}{2\pi \alpha} \ge 1$ and let $\delta^\aa=u-u^\aa$, then
\begin{equation*}
\hskip-.8in
\underset{0\le t\le T}{\sup}|\dd^\aa(t)|^2\le \tilde{ \epsilon}^2,
\end{equation*}
where $\tilde{\epsilon}$ is given in Theorem \ref{NSa-error-theorem}.
\end{theorem}
\begin{proof}
Taking the inner product of the above equation (\ref{MLa-delta1})  with $\dd^\aa$ we obtain
\begin{equation}
\hskip-.8in
\frac{1}{2}\frac{d}{dt}|\dd^\aa|^2+\nu \|\dd^\aa\|^2 \le |J_1|+|J_3|+|J_4|+|J_6|,  \label{MLa-error4}
\end{equation}
where $J_1,J_3$ and $J_4$ are as in (\ref{J-1}), (\ref{J-3}) and (\ref{J-4}) respectively and
\begin{equation*}
\hskip-.8in
J_6=((I+\alpha^2A)^{-1}(B(u^\aa,u^\aa)-B(v^\aa,u^\aa)), \dd^\aa).
\end{equation*}
Here again the estimates for $J_1,J_3$ and $J_4$ are as in the proof of Theorem \ref{LA-theorem}, with $\tilde{K}_0$ replaced by $\tilde{\tilde{K}}_{02}$ (see also Remark 6). Next, we estimate $J_6$. Thanks to (\ref{b-infty3}) we have
\begin{eqnarray*}
\hskip-.8in
|J_6|
&\le& |\alpha^2 ((I+\alpha^2A)^{-1}B(Au^\aa,u^\aa),\dd^\aa)|  \\
&=& |\alpha^2 (B(Au^\aa, (I+\alpha^2A)^{-1}\dd^\aa),u^\aa)| \\
&\le& c\alpha^2 \|u^\aa\|_{L^{\infty}} \|(I+\alpha^2 A)^{-1} \dd^\aa \||Au^\aa| \\
&\le& c\alpha^2 \|u^\aa\|_{L^{\infty}}|Au^\aa| \|\dd^\aa\| \\
&\le& \frac{\nu}{8}\|\dd^\aa\|^2+\frac{c\alpha^4}{\nu}|Au^\aa|^2\|u^\aa\|_{L^{\infty}}^2.
\end{eqnarray*}
Now since $\alpha^2|Au^\aa|^2 \le \tilde{\tilde{K}}_{02}^2$ by Proposition \ref{MLa-estimate}, then by Proposition \ref{infinity-inequality} (see also Remark 6) we have
\begin{equation}
\hskip-.8in
|J_6| \le \frac{\nu}{8}\|\dd^\aa\|^2 +\frac{c\alpha^4}{\nu} \tilde{\tilde{K}}_{02}^4 \left( 1+\log\left(\frac{L}{2\pi \alpha}\right) \right).  \label{J-6-estimate}
\end{equation}
From (\ref{J-1-estimate}), (\ref{J-3-estimate}), (\ref{J-4-estimate}) (where $\tilde{K}_0$ is replaced by
$\tilde{\tilde{K}}_{02}$), (\ref{J-6-estimate}) and (\ref{MLa-error4}) we have
\begin{equation*}
\hskip-.8in
\frac{d}{dt}|\dd^\aa|^2+\nu \|\dd^\aa\|^2 \le \frac{c}{\nu}\|u^\aa\|^2 |\dd^\aa|^2+\frac{c}{\nu}\alpha^2
\left( |f|^2+\tilde{\tilde{K}}_{02}^4\left( 1+\log \left( \frac{L}{2\pi \alpha} \right) \right) \right),
\end{equation*}
which concludes the proof of the theorem by Gronwall's inequality.
\end{proof}
\subsection{The rate of convergence of the simplified Bardina model}\label{subsecBardinaconv}
Now we present the error estimates for the simplified Bardina regularization model. As before, we will
proceed by estimating the $L^2$-norm of the difference $\delta^\aa=u-u^\aa$, where $u$ is the solution of 2D NSE system (\ref{NSE})
and $u^\aa$ is the solution of the simplified Bardina system (\ref{Bardina}).\\\\
From (\ref{NSE}) and (\ref{Bardina}) we observe that $\delta^\aa=u-u^\aa$ satisfies the following
equation:
\begin{eqnarray}
\hskip-.4in
\frac{d}{dt} \dd^\aa
+\nu A\dd^\aa+B(u,\dd^\aa)+B(\dd^\aa,u)-B(\dd^\aa,\dd^\aa)
&+&(I-(I+\alpha^2A)^{-1})B(u^\aa,u^\aa)\nonumber \\
&=&
(I-(I+\alpha^2A)^{-1})f.  \label{Bardina-delta1}
\end{eqnarray}
\begin{theorem}\label{Bardina-errortheorem}
Let $u^\aa$ be the solution of the 2D simplified Bardina system (\ref{Bardina}) with the initial date $u_0\in D(A)$, and let $u$ be the solution of the 2D NSE system (\ref{NSE}) with the same initial data $u_0$ over the interval $[0,T]$. Assume that
$\alpha$ is small enough such that $\frac{L}{2\pi \alpha} \ge 1$ and let $\delta^\aa=u-u^\aa$, then
\begin{equation*}
\hskip-.8in
\underset{0\le t \le T} {\sup} |\dd^\aa(t) |^2 \le \epsilon ^2,
\end{equation*}
where $\epsilon$ is given in Theorem \ref{LA-theorem}.
\end{theorem}
\begin{proof}
Taking the inner product of the equation (\ref{Bardina-delta1}) with $\dd^\aa$ we obtain
\begin{equation}
\hskip-.8in
\frac{1}{2}\frac{d}{dt}|\dd^\aa|^2+\nu \|\dd^\aa\|^2 \le |J_1|+|J_3|+|J_4|, \label{Bardina-error1}
\end{equation}
where $J_1,J_3$ and $J_4$ are as in (\ref{J-1}), (\ref{J-3}) and (\ref{J-4}) respectively. Here again
the estimates for $J_1,J_3$ and $J_4$ are as in the proof of Theorem \ref{LA-theorem}, with exactly the same $\tilde{K}_0$.\\
From (\ref{J-1-estimate}), (\ref{J-3-estimate}), (\ref{J-4-estimate}) and (\ref{Bardina-error1}) we have
\begin{equation*}
\hskip-.4in
\frac{d}{dt} |\dd^\aa|^2 +\nu \|\dd^\aa\|^2 \le \frac{c}{\nu} \|u\|^2 |\dd^\aa|^2 +\frac{c}{\nu}\alpha^2
\left( |f|^2+ \tilde{K}_0^4 \left(1+\log\left(\frac{L}{2\pi\alpha} \right) \right) \right).
\end{equation*}
Dropping the $\nu \|\dd^\aa\|^2$ term from the left-hand side and applying Gronwall's inequality we obtain
\begin{eqnarray*}
\hskip-.2in
|\dd^\aa(t)|^2
&\le& e^{\frac{c}{\nu}\int_0^t\|u(\ss)\|^2 d\ss} |\dd^\aa(0)|^2 \\
&\quad&+\frac{c\alpha^2}{\nu} \int_0^t
e^{\frac{c}{\nu} \int_\ss^t \|u(s)\|^2ds} \left(|f(\ss)|^2+\tilde{K}_0^4 \left(1+\log\left( \frac{L}{2\pi \alpha}\right) \right)\right) \: d\ss.
\end{eqnarray*}
Since $\dd^\aa(0)=0$, then by Proposition \ref{NSE-estimate} we obtain
\begin{equation*}
\hskip-.2in
|\dd^\aa(t)|^2
\le \frac{c\alpha^2}{\nu} e^{\frac{cK_0^2}{\nu^2}} \left( T\tilde{K}_0^4\left(1+\log \left( \frac{L}{2\pi\alpha} \right) \right)+\|f\|_{L^2((0,T);H)}^2 \right)=:\epsilon^2.
\end{equation*}
\end{proof}
\section{Error Estimates of The Galerkin Approximation of the $\alpha$-Regularization Models}\label{errorGalerkin}
In numerical simulation one needs to approximate the exact solutions of the underlying equations, that lie in the infinite dimensional function spaces, by functions that lie in finite dimensional spaces. In this section we estimate the errors between the approximation
solutions $u^\aa_m$, of the finite dimensional ordinary differential equation system (the finite dimensional Galerkin system of order $m$ in this context) and the
exact solutions $u^\aa$ of the $\alpha$-models in $L^2$-norm. The errors are given in terms of
$m$ and the regularization parameter $\aa$. Combining this with the results we establish in the previous section and by applying the triangle inequality, we obtain error estimates of numerical approximation solutions
$u_m^\aa$ of the underlying $\aa$-model and of the exact solution $u$ of the 2D NSE system. We take the Leray-$\alpha$ model
as an example in this section and show the error estimate results. By similar arguments,  one can show the error estimates for the other $\alpha$-regularization models introduced in this paper.\\\\
\noindent
{\bf{Galerkin approximation for the 2D Leray-$\alpha$ model}}\\
\noindent
Now we present the error estimates for the Galerkin approximation of the Leray-$\alpha$ model.
This estimate for the rate of convergence of the Leray-$\alpha$ model is along the same lines of
\cite{DMT, Heywood2, Rautmann, Rautmann2} and in the spirit of the work for the 2D NSE.\\\\
For $ u^\alpha$ an exact solution of the Leray-$\alpha$ system (\ref{LA}), we can decompose $u^\aa$ as follows: $u^\alpha=p_m+q_m$, where $p_m=P_m\: u^\aa$ and $q_m=(I-P_m)u^\aa$, $P_m$ is the orthogonal project from $H$ onto $H_m$, which is defined in Section \ref{sec-fun}. Rewriting equation (\ref{LA}) as
\begin{equation}
\hskip-.8in
\frac{du^\aa}{dt}+\nu Au^\aa+(I+\alpha^2A)^{-1}B(u^\aa, (I+\alpha^2A)u^\aa)=(I+\alpha^2A)^{-1}f,
\label{LAnew}
\end{equation}
and since $u^\aa=p_m+q_m$, we can decompose
the above equation (\ref{LAnew}) into the following coupled system of equations:
\begin{eqnarray}
\hskip-1.5in
&&\frac{d p_m}{dt}+\nu Ap_m +(I+\alpha^2A)^{-1} P_m\: B(u^\alpha, (I+\alpha^2A)u^\alpha)=(I+\alpha^2A)^{-1}P_mf, \label{LA-lowmode} \\
&&
\frac{dq_m}{dt}+\nu Aq_m+(I+\alpha^2A)^{-1}(I-P_m) B (u^\alpha,(I+\alpha^2A)u^\alpha)=(I+\alpha^2A)^{-1}(I-P_m)f. \label{LA-highmode}
\end{eqnarray}
For the Galerkin approximation system of the Leray-$\alpha$ system, we rewrite the equation (\ref{LAGalerkin}) in the following equivalent form
\begin{equation}
\hskip-.8in
\frac{du_m^\aa}{dt}+\nu Au_m^\aa+(I+\alpha^2A)^{-1}P_m B(u_m^\aa, (I+\alpha^2A)u_m^\aa)=(I+\alpha^2A)^{-1}P_mf. \label{LAGalerkinnew}
\end{equation}\\
We will proceed by first estimating the $L^2$-norm of $q_m$ and then the $L^2$-norm of the difference $\dd_m=p_m-u_m^\aa$, where $u_m^\aa$ is the
solution of the Galerkin system of the Leray-$\alpha$ system (\ref{LAGalerkinnew}). Then by the triangle inequality and orthogonality of spaces projected by $P_m$ and $(I-P_m),$ we obtain the error estimates of
the $L^2$-norm: $|u^\aa-u_m^\aa|^2=|q_m|^2+|p_m-u_m^\aa|^2$. \\\\
From (\ref{LA-lowmode}) and (\ref{LAGalerkinnew}) we observe that $\delta_m=p_m-u_m^\aa$ satisfies the following equation:
\begin{equation*}
\frac{d\dd_m}{dt} +\nu A\dd_m+(I+\alpha^2A)^{-1}P_mB(u^\aa,(I+\alpha^2A)u^\aa)
-(I+\alpha^2A)^{-1}P_mB(u_m^\aa,(I+\alpha^2A)u_m^\aa)=0.
\end{equation*}
Since $u^\aa=p_m+q_m$ and $\dd_m=p_m-u_m^\aa$, we can rewrite the above equation as
\begin{eqnarray}
\hskip-.8in
\frac{d \dd_m}{dt}+\nu A\dd_m
&+&(I+\alpha^2A)^{-1}P_mB(\dd_m+q_m,(I+\alpha^2A)u^\aa)   \nonumber \\
&-&
(I+\alpha^2A)^{-1}P_mB(u_m^\aa,(I+\alpha^2A)(\dd_m+q_m))=0.\label{LA-delta}
\end{eqnarray}
\begin{theorem} \label{LA-Galerkinerror}
Let $T>0$ and let $u^\aa$ be a solution of the Leray-$\alpha$ system (\ref{LAnew}) with initial data $u_0\in D(A)$, and let
$u_m^\aa$ be the solution of (\ref{LAGalerkinnew}) with initial data $u_{0m}=P_m \:u_0$ over the interval $[0,T]$.
For a given $m\ge 1$,  assume that $\alpha$ is small such that $\aa^2\le\frac{1}{\lambda_{m+1}}$, then
\begin{equation*}
\hskip-.8in
\underset{0\le t\le T} {\sup}|u^\aa(t)-u_m^\aa(t)|^2 \le e ^2,
\end{equation*}
where
$e^2:=\frac{1}{\lambda_{m+1}^2} (Q+R+L_m\tilde{U}\tilde{V})$, $L_m=1+\log(\frac{\lambda_m}{\lambda_1})$ and $Q,R,\tilde{U}, \tilde{V}$ depend on $\nu, u_0, \log(\frac{L}{2\pi\alpha}), f, T$, and are given explicitly in (\ref{Q}), (\ref{R}), (\ref{tildeU}) and (\ref{tildeV}), respectively.
\end{theorem}
\begin{remark}
Here we require $f \in L^\infty((0,T); H)$, which is stronger assumption than the condition $f\in L^2((0,T);H)$ in the
estimate of rates of convergence in section \ref{errorconv}.
\end{remark}
\begin{proof}
First, we estimate the $L^2$ norm of $q_m$. \\
We take the inner product of equation (\ref{LA-highmode}) with $q_m$ and get
\begin{eqnarray}
\hskip-.0in
\frac{1}{2}\frac{d}{dt}|q_m|^2+\nu \|q_m\|^2
& \le& | ( (I+\alpha^2A)^{-1}B(u^\alpha, (I+\alpha^2A)u^\aa),q_m)| \nonumber \\
&\quad &
+| ( (I+\alpha^2A)^{-1}f, q_m)| \nonumber \\
&\le & M_1+M_2.   \label{qinnerproduct}
\end{eqnarray}
Next, we estimat $M_1,\:M_2$. By virtue of (\ref{b-infty}) we have
\begin{eqnarray*}
\hskip-.8in
M_1
&= &| ( (I+\alpha^2A)^{-1}B(u^\alpha, (I+\alpha^2A)u^\aa),q_m)| \\
&=& |(B(u^\aa,(I+\alpha^2A)u^\aa), (I+\alpha^2A)^{-1}q_m)| \\
&\le&c \|u^\aa\|_{L^\infty} \|(I+\alpha^2A)u^\aa\||(I+\alpha^2A)^{-1}q_m| \\
&\le& c\tilde{K}_0(1+\log(\frac{L}{2\pi \alpha}) )^{1/2}(\|u^\aa\|+\alpha^2 |A^{3/2}u^\aa|) \frac{\|q_m\|}{\lambda_{m+1}^{1/2}}\:,
\end{eqnarray*}
where in the last inequality we apply (\ref{pc1}) and Proposition \ref{infinity-inequality} to $\|u^\aa\|_{L^\infty}$. Notice
that by the assumption that $\alpha^2\le \frac{1}{\lambda_{m+1}} \le \frac{1}{\lambda_1}=(\frac{L}{2\pi})^2$, we consequently have $\frac{L}{2 \pi \alpha}\ge 1$, and as a result, it is valid to apply Proposition \ref{infinity-inequality}.\\\\
Now, by Young's inequality and the \textit{a priori} estimates obtained in (\ref{LA-apriori}), we have
\begin{eqnarray}
\hskip-.8in
M_1& \le& \frac{\nu}{4} \|q_m\|^2+ \frac{1}{\lambda_{m+1}} \frac{c}{\nu}\tilde{K}_0^2 (1+\log(\frac{L}{2\pi\alpha}))(\|u^\aa\|^2+\alpha^4 |A^{3/2}u^\aa|^2)  \nonumber \\
&\le&\frac{\nu}{4} \|q_m\|^2+ \frac{1}{\lambda_{m+1}} \frac{c}{\nu}\tilde{K}_0^4 (1+\log(\frac{L}{2\pi\alpha})) \\
&\quad& +\frac{\alpha^2}{\lambda_{m+1}}\frac{c}{\nu} \tilde{K}_0^2(1+\log(\frac{L}{2\pi\alpha}))(\alpha^2|A^{3/2}u^\aa|^2) .
\label{M1-result}
\end{eqnarray}
Now, for estimating $M_2$, we apply Cauchy-Schwarz inequality, (\ref{pc1}) and Young's inequality to obtain
\begin{eqnarray}
\hskip-.8in
M_2
&=&|( (I+\alpha^2A)^{-1}(I-P_m)f, q_m)|=|((I-P_m)f, (I+\alpha^2A)^{-1}q_m)| \nonumber \\
&\le& |f||(I+\alpha^2A)^{-1}q_m|
\le |f| \frac{\|q_m\|}{\lambda_{m+1}^{1/2}} \le \frac{\nu}{4}\|q_m\|^2+\frac{1}{\lambda_{m+1}}\frac{c}{\nu}|f|^2. \label{M2-result}
\end{eqnarray}
Let us substitute the bounds for $M_1,\: M_2$ into (\ref{qinnerproduct}) to obtain
\begin{eqnarray}
\hskip-.4in
\frac{d}{dt}|q_m|^2+ \nu \|q_m\|^2
&\le& \frac{1}{\lambda_{m+1}}\frac{c}{\nu}\left( |f|^2+\tilde{K}_0^4 (1+\log(\frac{L}{2\pi\alpha})) \right)\nonumber \\
&\quad &
+\frac{\alpha^2}{\lambda_{m+1}}\frac{c}{\nu}\tilde{K}_0^2(1+\log(\frac{L}{2\pi\alpha}))(\alpha^2 |A^{3/2}u^\aa|^2). \label{q-equation}
\end{eqnarray}
By Poincar\'{e} inequality and (\ref{pc1}), $\lambda_{m+1} |q_m|^2\le \|q_m\|^2$, we obtain
\begin{eqnarray*}
\hskip-.4in
\frac{d}{dt}|q_m|^2+ \nu\lambda_{m+1} |q_m|^2
&\le& \frac{1}{\lambda_{m+1}}\frac{c}{\nu}\left( |f|^2+\tilde{K}_0^4 (1+\log(\frac{L}{2\pi\alpha})) \right)\\
&\quad &
+\frac{\alpha^2}{\lambda_{m+1}}\frac{c}{\nu}\tilde{K}_0^2(1+\log(\frac{L}{2\pi\alpha}))(\alpha^2 |A^{3/2}u^\aa|^2).
\end{eqnarray*}
By Gronwall inequality, we get
\begin{eqnarray*}
|q_m(t)|^2
&\le& e^{-\nu \lambda_{m+1}t} |q_m(0)|^2
+\int_0^t e^{-\nu \lambda_{m+1}(t-s)}\frac{1}{\lambda_{m+1}}\frac{c}{\nu} \left(|f|^2+\tilde{K}_0^4(1+\log(\frac{L}{2\pi\alpha}))\right)\: ds \nonumber  \\
&\quad&+ \int_0^t e^{-\nu \lambda_{m+1}(t-s)} \frac{\alpha^2}{\lambda_{m+1}} \frac{c}{\nu}\tilde{K}_0^2(1+\log(\frac{L}{2\pi\alpha}))
(\alpha^2|A^{3/2}u^\aa|^2)\:ds       \nonumber \\
&\le& e^{-\nu \lambda_{m+1}t} \frac{|Aq_m(0)|^2}{\lambda_{m+1}^2}
      +\frac{1}{\lambda_{m+1}^2} (1-e^{-\nu\lambda_{m+1} t} )\frac{c}{\nu^2}\left(|f|_{L^\infty((0,T);H)}^2+ \tilde{K}_0^4(1+\log(\frac{L}{2\pi\alpha})) \right)      \nonumber    \\
     &\quad& + \frac{\alpha^2}{\lambda_{m+1}} \frac{c}{\nu}\tilde{K}_0^2(1+\log(\frac{L}{2\pi\alpha})) \int_0^t (\alpha^2|A^{3/2}u^\aa|^2) \:ds      \nonumber   \\
     &\le& \frac{1}{\lambda_{m+1}^2}|Au_0|^2 +\frac{1}{\lambda_{m+1}^2} \frac{c}{\nu^2}\left(|f|_{L^\infty((0,T):H)}^2+\tilde{K}_0^4(1+\log(\frac{L}{2\pi\alpha}))\right)     \nonumber    \\
     &\quad&+\frac{\alpha^2}{\lambda_{m+1}}\frac{c}{\nu^2}\tilde{K}_0^4
     (1+\log(\frac{L}{2\pi\alpha})).  \nonumber
\end{eqnarray*}
where we apply the \textit{a priori} estimates of solutions of Leray-$\alpha$ model given in Proposition
\ref{LA-estimate}. Next, denote by
\begin{equation}
Q:=|Au_0|^2+\frac{c}{\nu^2}\left(|f|_{L^\infty((0,T):H)}^2+\tilde{K}_0^4(1+\log(\frac{L}{2\pi\alpha}))\right),\label{Q}
\end{equation}
and
\begin{equation}
R:=\frac{c}{\nu^2}\tilde{K}_0^4(1+\log(\frac{L}{2\pi\alpha})). \label{R}
\end{equation}
we obtain
\begin{equation*}
\hskip-.8in
|q_m(t)|^2\le \frac{1}{\lambda_{m+1}^2}Q+\frac{\alpha^2}{\lambda_{m+1}}R.
\end{equation*}
By the assumption that $\alpha^2\le \frac{1}{\lambda_{m+1}}$, we have
\begin{equation}
\hskip-.8in
|q_m|^2\le \frac{1}{\lambda_{m+1}^2}(Q+R). \label{q-result}
\end{equation}\\\\
Next, we estimate the $L^2$-norm of $\delta_m$.\\
Taking the inner product of equation (\ref{LA-delta}) with $\dd_m$, we get
\begin{equation}
\hskip-.8in
\frac{1}{2}\frac{d}{dt} |\dd_m|^2+\nu \|\dd_m\|^2 \le M_3+M_4+M_5+M_6, \label{dinnerproduct}
\end{equation}
where
\begin{eqnarray*}
\hskip-1in
&&M_3=|B(\dd_m,(I+\alpha^2A)u^\aa),(I+\alpha^2A)^{-1}\dd_m)|, \\
\hskip-1in
&& M_4=|B(q_m,(I+\alpha^2A)u^\aa),(I+\alpha^2A)^{-1}\dd_m)|,\\
\hskip-1in
&& M_5=|(B(u_m^\aa,(I+\alpha^2A)\dd_m), (I+\alpha^2A)^{-1}\dd_m)|,\\
\hskip-1in
&&M_6=|(B(u_m^\aa,(I+\alpha^2A)q_m),(I+\alpha^2A)^{-1}\dd_m)|.
\end{eqnarray*}
First, let us estimate $M_3$. Thanks to (\ref{b-uvw}) we have
\begin{eqnarray}
M_3
&\le& c|\dd_m|^{1/2}\|\dd_m\|^{1/2}\|(I+\alpha^2A)u^\aa\|  |(I+\alpha^2A)^{-1}\dd_m|^{1/2}\|(I+\alpha^2A)^{-1}\dd_m\|^{1/2}\nonumber \\
&\le& c|\dd_m| \|\dd_m\| ( \|u^\aa\|+\alpha^2|A^{3/2}u^\aa|)\nonumber \\
&\le& \frac{\nu}{16}\|\dd_m\|^2+\frac{c}{\nu}|\dd_m|^2(\|u^\aa\|^2+\alpha^4|A^{3/2}u^\aa|^2) \nonumber \\
&\le& \frac{\nu}{16}\|\dd_m\|^2+\frac{c}{\nu}|\dd_m|^2\tilde{K}_0^2 +\frac{1}{\lambda_{m+1}}\frac{c}{\nu}|\dd_m|^2(\aa^2|A^{3/2}u^\aa|^2), \label{M3-result}
\end{eqnarray}
where in the last inequality we use the \textit{a priori} estimates of Leray-$\alpha$ obtained in Proposition
\ref{LA-estimate} and the assumption that $\alpha^2 \le \frac{1}{\lambda_{m+1}}$.\\
Now, we estimate $M_4$. By applying (\ref{b-bg2}) and (\ref{log}) we obtain
\begin{eqnarray}
M_4
&\le&c|q_m|\|(I+\aa^2A)u^\aa\| \|(I+\aa^2A)^{-1}\dd_m\|_{L^\infty} \nonumber \\
&\le&c|q_m|(\|u^\aa\|+\alpha^2|A^{3/2}u^\alpha|)\|(I+\alpha^2A)^{-1}\dd_m\|\left(1+\log( \frac{\lambda_m}{\lambda_1})\right)^{1/2}\nonumber \\
&\le & c|q_m|(\|u^\aa\|+\alpha^2|A^{3/2}u^\aa|) \|\dd_m\|L_m^{1/2}\nonumber \\
&\le&\frac{\nu}{16}\|\dd_m\|^2+\frac{cL_m}{\nu}|q_m|^2(\|u^\aa\|^2+\alpha^2(\alpha^2|A^{3/2}u^\aa|^2))\nonumber\\
&\le& \frac{\nu}{16}\|\dd_m\|^2+\frac{cL_m}{\nu}|q_m|^2(\tilde{K}_0^2+\alpha^2(\alpha^2|A^{3/2}u^\aa|^2)),\label{M4-result}
\end{eqnarray}
where $L_m=1+\log(\frac{\lambda_m}{\lambda_1})$, and we apply Proposition \ref{BG-inequality} and (\ref{log}) in the second inequality above. \\
Let us now estimate $M_5$. By virtue of (\ref{b-uvw}) we have
\begin{eqnarray*}
\hskip-.8in
M_5
&\le& c|u_m^\aa|^{1/2}\|u_m^\aa\|^{1/2}\|(I+\aa^2A)\dd_m\| |(I+\aa^2A)^{-1}\dd_m|^{1/2} \|(I+\aa^2A)^{-1}\dd_m\|^{1/2} \\
&\le&c|u_m^\aa|^{1/2}\|u_m^\aa\|^{1/2}(1+\alpha^2 \lambda_m)|\|\dd_m\||\dd_m|^{1/2}\|\dd_m\|^{1/2}.
\end{eqnarray*}
Since $\alpha^2\lambda_m\le\alpha^2 \lambda_{m+1}\le 1$, and $c$ denotes general dimensionless constant, we have
\begin{eqnarray}
\hskip-.8in
M_5
&\le& c\|\dd_m\|^{3/2}|\dd_m|^{1/2}|u_m^\aa|^{1/2}\|u_m^\aa\|^{1/2} \nonumber \\
&\le& \frac{\nu}{16}\|\dd_m\|^2+\frac{c}{\nu}|\dd_m|^2 |u_m^\aa|^2\|u_m^\aa\|^2\nonumber\\
&\le&\frac{\nu}{16}\|\dd_m\|^2+\frac{c}{\nu^3}|\dd_m|^2 \frac{\tilde{K}_0^4}{\lambda_1}, \label{M5-result}
\end{eqnarray}
where in the last inequality we apply Proposition \ref{LA-estimate}.\\
For the last term $M_6$, we will proceed by applying (\ref{bilinear}) and (\ref{b-infty}), then we obtain
\begin{eqnarray*}
M_6
&=&|(B(u_m^\aa,(I+\aa^2A)q_m),(I+\aa^2A)^{-1}\dd_m)| \nonumber \\
&=& |(B(u_m^\aa,(I+\aa^2A)^{-1}\dd_m), (I+\aa^2A)q_m)| \nonumber \\
&\le &c \|u_m^\aa\|_{L^\infty}\|(I+\aa^2A)^{-1}\dd_m\||(I+\aa^2A)q_m|.     \nonumber
\end{eqnarray*}
By applying Proposition \ref{BG-inequality}, inequality (\ref{log}), (\ref{pc1}), and the facts that
\begin{eqnarray*}
&&\|(I+\alpha^2A)^{-1} \phi \| \le \| \phi\|, \quad \mbox{ for all } \phi \in V, \\
&&|(I+\alpha^2A)\phi| \sim |\phi|+\alpha^2|A\phi| \quad \mbox{ for all } \phi \in D(A),
\end{eqnarray*}
we obtain
\begin{eqnarray}
M_6
&\le& c\|u_m^\aa\| \left(1+\log(\frac{\lambda_m}{\lambda_1})\right)^{1/2}\| (I+\alpha^2A)^{-1}\dd_m\| |(I+\alpha^2A)qm| \nonumber \\
&\le& cL_m^{1/2}\tilde{K}_0\|\dd_m\|(|q_m|+\alpha^2 |Aq_m|) \nonumber   \\
&\le&cL_m^{1/2}\tilde{K}_0 \|\dd_m\| (|q_m|+\aa^2 \frac{|A^{3/2}q_m|}{\lambda_{m+1}^{1/2}}) \nonumber \\
&\le& \frac{\nu}{16}\|\dd_m\|^2+\frac{cL_m}{\nu}\tilde{K}_0^2(|q_m|^2+\frac{\alpha^2}{\lambda_{m+1}}(\alpha^2|A^{3/2}u^\aa|^2))   \nonumber\\
&\le& \frac{\nu}{16}\|\dd_m\|^2+\frac{cL_m}{\nu}\tilde{K}_0^2|q_m|^2+\frac{1}{\lambda_{m+1}^2}
\frac{cL_m}{\nu}\tilde{K}_0^2(\aa^2|A^{3/2}u^\aa|^2),\label{M6-result}
\end{eqnarray}
where $L_m=1+\log(\frac{\lambda_m}{\lambda_1})$, and in the second inequality above we use the fact that
$\|u_m^\aa\|\le \tilde{K}_0$ from Proposition \ref{LA-estimate}.\\\\
Plugging (\ref{M3-result}), (\ref{M4-result}), (\ref{M5-result}) and (\ref{M6-result}) into the inequality
(\ref{dinnerproduct}) we obtain
\begin{eqnarray*}
\frac{d}{dt}|\dd_m|^2+\nu \|\dd_m\|^2
&\le& \frac{c}{\nu}|\dd_m|^2(\tilde{K}_0^2 +\frac{1}{\lambda_{m+1}}(\aa^2|A^{3/2}u^\aa|^2)+\frac{\tilde{K}_0^4}{\nu^2\lambda_1})\\
&\quad& +\frac{cL_m}{\nu}\tilde{K}_0^2|q_m|^2
+\alpha^2\frac{cL_m}{\nu}|q_m|^2(\alpha^2|A^{3/2}u^\aa|^2)
\nonumber\\
&\quad&+\frac{1}{\lambda_{m+1}^2}\frac{cL_m}{\nu}\tilde{K}_0^2(\aa^2|A^{3/2}u^\aa|^2) \nonumber \\
&\le& |\dd_m|^2 U+ L_mV,
\end{eqnarray*}
where
\begin{equation}
U(t):=\frac{c}{\nu}(\tilde{K}_0^2+\frac{1}{\lambda_{m+1}}(\aa^2|A^{3/2}u^\aa|^2)+\frac{\tilde{K}_0^4}{\nu^2\lambda_1}),\label{U}
\end{equation}
and
\begin{equation}
V(t):= \frac{c}{\nu}\tilde{K}_0^2|q_m|^2
+\alpha^2\frac{c}{\nu}|q_m|^2(\alpha^2|A^{3/2}u^\aa|^2)
+\frac{1}{\lambda_{m+1}^2}\frac{c}{\nu}\tilde{K}_0^2(\aa^2|A^{3/2}u^\aa|^2). \label{V}
\end{equation}
Applying Gronwall inequality and recalling that $|\dd_m(0)|=0$, we obtain
\begin{equation}
|\dd_m(t)|^2
\le L_m\int_0^T e^{\int_\tau^t U(s)ds}V(\tau)\:d\tau,\: t\in [0,T]. \label{Gronwall}
\end{equation}
By the \textit{a priori} estimates in Proposition \ref{LA-estimate}, we know that
\begin{equation}
\nu \int_0^T (\aa^2|A^{3/2}u^\aa|^2)\: dt \le \tilde{K}_0^2,\label{refer}
\end{equation}
which implies that $U\in L^1(0,T)$ and consequently we have
\begin{equation}
\int_0^T U(t)\:dt \le \frac{c}{\nu}(\tilde{K}_0^2T+\frac{1}{\nu\lambda_{m+1}}\tilde{K}_0^2+\frac{\tilde{K}_0^4}{\nu^2\lambda_1}T)=:\tilde{U}.
\label{tildeU}
\end{equation}
Now for estimating $V(t)$,  we recall from (\ref{q-result}) that $|q_m|^2\le \frac{1}{\lambda_{m+1}^2}(Q+R)$. Applying
(\ref{refer}), we obtain
\begin{equation}
\int_0^T V(t)\: dt \le \frac{1}{\lambda_{m+1}^2}\left( \frac{c}{\nu}\tilde{K}_0^2(Q+R)T+\frac{1}{\lambda_{m+1}}\frac{c}{\nu^2}\tilde{K}_0^2(Q+R)+\frac{c}{\nu^2}\tilde{K}_0^4\right)=:\frac{1}{\lambda_{m+1}^2}\tilde{V}. \label{tildeV}
\end{equation}
Plugging back into the inequality (\ref{Gronwall}), we get
\begin{equation}
\hskip-.8in
|\dd_m(t)|^2\le\frac{L_m}{\lambda_{m+1}^2}\tilde{U}\tilde{V} .\label{deltaportion}
\end{equation}
Now by Pythagorean Theorem and orthogonality of the projection spaces $P_m$ and $(I-P_m)$, we have
\begin{equation}
\underset{0\le t\le T}{\sup}| u^\aa(t)-u_m^\aa(t)|^2 \le \underset{0\le t\le T}{\sup} |\dd_m|^2+\underset{0\le t\le T}{\sup} |q_m|^2 \le \frac{1}{\lambda_{m+1}^2} (Q+R+L_m\tilde{U}\tilde{V}), \label{resultresult}
\end{equation}
where $Q,R$ and $\tilde{U}, \tilde{V}$ are given in (\ref{Q}), (\ref{R}), (\ref{tildeU}) and (\ref{tildeV}), respectively.
\end{proof}
\begin{remark}
The above result $|u^\aa(t)-u^\aa_m(t)|=O\left( \frac{1}{\lambda_mL^2}(\log(\lambda_mL^2))^{1/2}\right)$
is of the same order as that of the error estimates for the usual Galerkin approximation of NSE.
Indeed, in \cite{Titi} the author points out that for the 2D NSE, the error estimate for the usual Galerkin,
$|u-u_m|$ is of order $O\left(\frac{1}{\lambda_mL^2}(\log(\lambda_mL^2))^{1/2})\right)$, where $u$ is the solution of
2D NSE (\ref{NSE}) and $u_m$ is the solution of the corresponding Galerkin system (\ref{NSEGal}). Furthermore, this estimate is optimal up to the logarithmic terms provided $f\in L^2$.
\end{remark}
Now, simply by applying the triangle inequality, we achieve the error estimates of the solution $u^\aa_m$ for the finite-dimensional Galerkin system of the 2D Leray-$\alpha$ model as approximation of the exact solution
$u$ of the 2D NSE system.
\begin{theorem}\label{finalthm}
Let $T>0$ and let $u$ be the solution of the 2D NSE system (\ref{NSE}) with initial data $u_0\in D(A)$ and $u^\aa_m$ be the solution of finite-dimensional Galerkin approximation of the 2D Leray-$\alpha$ system (\ref{LAGalerkinnew}) with initial data $u_{0m}=P_m\:u_0$ over the interval $[0,T]$. For a given $m\ge 1$, assume that $\alpha$ is small enough such that $\alpha \le \frac{2\pi}{\lambda_{m+1}L}$, where $\lambda_{m+1}$ is the $(m+1)$-th eigenvalue of the Stokes operator $A$, then
\begin{equation}
\hskip-.8in
\underset{0\le t\le T}{\sup} |u(t)-u^\aa_m(t)|^2 \le C \left( (\frac{(2\pi)^2}{\lambda_{m+1}L^2})^2\log\left(\frac{\lambda_{m+1}L^2}{(2\pi)^2}\right)\right),
\end{equation}
where $C$ is a constant, which depends on $\nu, u_0, f, L$ and $T$ only.
\end{theorem}
\begin{proof}
By the triangle inequality we have
\begin{equation}
\hskip-.8in
|u(t)-u^\aa_m(t)|^2 \le 2(|u(t)-u^\aa(t)|^2+|u^\aa(t)-u^\aa_m(t)|^2). \label{finaltheorem1}
\end{equation}
From Theorem \ref{LA-theorem} we obtain, under the assumption that $\alpha\le \frac{L}{2\pi}$,
\begin{equation}
\hskip-.8in
\underset{0\le t\le T} {\sup}|u(t)-u^\aa(t)|^2 \le C' \left( \left(\frac{2\pi\alpha}{L}\right)^2\log\left(\frac{L}{2\pi\alpha}\right)\right),\label{finaltheorem2}
\end{equation}
where $C'$ is a constant that depends only on $\nu, u_0, f, L$ and $T$.
By virtue of Theorem \ref{LA-Galerkinerror} above, we have, under the assumption that $\alpha^2\le \frac{1}{\lambda_{m+1}^2}$,
\begin{equation}
\hskip-.8in
\underset{0\le t\le T} {\sup}|u^\aa(t)-u^\aa_m(t)|^2 \le \tilde{C} \left( \left(\frac{(2\pi)^2}{\lambda_{m+1}L^2}\right)^2\log\left(\frac{\lambda_{m+1}L^2}{(2\pi)^2}\right)\right),\label{finaltheorem3}
\end{equation}
where $\tilde{C}$ is a constant that depends only on $\nu, u_0, f, L$ and $T$.\\
Now we assume that $\alpha\le \frac{2\pi}{\lambda_{m+1}L}$. Noticing that $\lambda_1=(\frac{2\pi}{L})^2$, we have
\begin{equation}
\hskip-.8in
\alpha^2\le (\frac{2\pi}{\lambda_{m+1}L})(\frac{2\pi}{\lambda_{m+1}L})=\frac{\lambda_1}{\lambda_{m+1}}\frac{1}{\lambda_{m+1}}\le \frac{1}{\lambda_{m+1}},
\end{equation}
where in the last step we use the fact that $\lambda_1\le \lambda_{m+1}$.\\
Since both assumptions for (\ref{finaltheorem2}) and (\ref{finaltheorem3}) are satisfied, we combine
(\ref{finaltheorem1}), (\ref{finaltheorem2}) and (\ref{finaltheorem3}) to obtain
\begin{equation*}
\hskip-.8in
\underset{0\le t\le T}{\sup} |u(t)-u^\aa_m(t)|^2\le C \left( \left(\frac{(2\pi)^2}{\lambda_{m+1}L^2}\right)^2\log\left(\frac{\lambda_{m+1}L^2}{(2\pi)^2}\right)\right),
\end{equation*}
which concludes our proof.
\end{proof}

Recall that, according to (\ref{asymptotic}), the $(m+1)$-th eigenvalue of the 2D Stokes operator satisfies
\begin{equation}
\hskip-.8in
\frac{m+1}{c_0}\le \frac{\lambda_{m+1}}{\lambda_1}\le c_0 (m+1),
\end{equation}
where $c_0$ is a constant that depends only on $L$. Applying the above asymptotic estimate and the fact $\lambda_1=(\frac{2\pi}{L})^2$, one can rewrite
the conditions and results of Theorem \ref{finalthm} in the form of the following corollary:
\begin{corollary}
Let $T>0$ and let $u$ be the solution of the 2D NSE system (\ref{NSE}) with initial data $u_0\in D(A)$ and $u^\aa_m$ be the solution of finite-dimensional Galerkin approximation of the 2D Leray-$\alpha$ system (\ref{LAGalerkinnew}) with initial data $u_{0m}=P_m\:u_0$ over the interval $[0,T]$. For a given $m\ge 1$, assume that $\alpha$ is small enough such that $\alpha \le \frac{L}{2\pi}\frac{c_0}{m+1}$, where
$c_0$ is a constant that depends only on $L$ as stated in (\ref{asymptotic}), then
\begin{equation}
\hskip-.8in
\underset{0\le t\le T}{\sup} |u(t)-u^\aa_m(t)|^2 \le C_0(\frac{1}{m+1})^2\log(m+1)
\end{equation}
where $C_0$ is a constant, which depends on $\nu, u_0, f, L$ and $T$ only.
\end{corollary}
The error estimates for the other $\alpha$-models are not presented in this paper in detail. The results are the same as that of the Leray-$\alpha$ model we present above. One can follow the same idea
and check the calculation. In fact, the proof of the estimates for the Modified Leray-$\alpha$ model and simplified Bardina model will follow readily from the proof of the Leray-$\alpha$ model since the
former two have milder nonlinearity and the estimates on the nonlinear terms will be easier
compared to that of the Leray-$\alpha$ model. For the 2D Navier-Stokes-$\alpha$ equation, combining
(\ref{b-infty})-(\ref{b-infty3}) and (\ref{tildeB-id}), we can get similar inequalities for $\tilde{B}$ as those of
$B$ and by following same steps obtain similar error estimates for the Navier-Stokes-$\alpha$ model.
In a subsequent paper we will present error estimates regarding the
three-dimensional case in the spirit of the results reported in \cite{CGKTW}.

\section*{appendix}
For the sake of completeness, we present in this section the version of the Brezis-Gallouet inequality
presented in Proposition \ref{BG-inequality}.
\begin{proof}
Let $\vp \in D(A)$ such that
\begin{equation*}
\hskip-.8in
\vp=\underset{k \in \Z^2\backslash   {(0,0)}}{\Sigma}\hat{\vp}_ke^{i2\pi \frac{x\cdot k}{L}}
\end{equation*}
therefore,
\begin{eqnarray*}
\hskip-.8in
\| \vp\|_{L^{\infty}}
&\le& \underset{k\in \Z^2 \backslash {(0,0)}} {\Sigma}| \hat{\vp}_k |
=\underset{0<|k|\le M}{\Sigma} | \hat{\vp}_k |+\underset{|k|>M}{\Sigma}| \hat{\vp}_k |  \\
&=& \underset{0<|k|\le M} {\Sigma}\frac{ (1+|k|^2)^{1/2}}{(1+|k|^2)^{1/2}}|\hat{\vp}_k| +
\underset{|k|>M}{\Sigma}\frac{(1+|k|^2)}{(1+|k|^2)}|\hat{\vp}_k|,
\end{eqnarray*}
where $M\ge 1$ is a real number to be determined later.\\
Then by the Cauchy-Schwarz inequality we have
\begin{eqnarray*}
\hskip-.8in
\|\vp\|_{L^{\infty}}
&\le& \left( \underset{0<|k|\le M}{\Sigma} (1+|k|^2)| \hat{\vp}_k|^2 \right)^{1/2}
\left( \underset{0<|k|\le M} {\Sigma} \frac{1}{(1+|k|^2)} \right)^{1/2} \\
&+&\left( \underset{|k|>M}{\Sigma} (1+|k|^2)^2 | \hat{\vp}_k|^2 \right)^{1/2}
\left(\underset{|k|>M}{\Sigma} \frac{1}{(1+|k|^2)^2} \right)^{1/2}    \\
&\le& c\|\vp \| \left(\int_{|y|\le M}\frac{dy}{1+|y|^2} \right)^{1/2} +
c\frac{L}{2\pi} |A\vp| \left( \int_{|y|\ge M} \frac{dy}{(1+|y|^2)^2 } \right)^{1/2}.
\end{eqnarray*}
We observe that in two dimensions
\begin{equation*}
\hskip-.8in
\int_{|y|\le M} \frac{dy}{1+|y|^2} = \int_0^{2\pi}d \theta \int_0^M\frac{r\: dr}{1+r^2}=\pi \log(1+M^2),
\end{equation*}
and
\begin{equation*}
\hskip-.8in
\int_{|y|\ge M} \frac{dy}{(1+|y|^2)^2}=\int_0^{2\pi} d\theta \int_M^{\infty} \frac{r \: dr}{(1+r^2)^2}=\pi\frac{1}{1+M^2}.
\end{equation*}
Therefore,
\begin{equation*}
\hskip-.8in
\|\vp\|_{L^{\infty}} \le
c\left( \|\vp\| \left(\log (1+M^2) \right)^{1/2}+\frac{L}{2\pi} \frac{|A\vp |}{(1+M^2)^{1/2} } \right)
\end{equation*}
for every $M\ge 1$. Notice that by Poincar\'{e} inequality we have $\frac{L}{2\pi} \frac{|A\vp|}{\|\vp\|} \ge 1$,
therefore, one can choose $M=\sqrt{ (\frac{L}{2\pi}\frac{|A\vp|}{\|\vp\|} )^2-1}+1 $ to conclude the proof.
\end{proof}
\section*{Acknowledgements}
This work was supported in part by the NSF grants no. DMS-0504619 and no. DMS-0708832, and by the ISF grant no. 120/06.

\end{document}